\documentstyle[prc,aps]{revtex}
\begin{document}\draft
\title{Effective nucleon-nucleon interactions and nuclear matter equation 
of state}
\author{K.C. Chung$^a$, C.S. Wang$^{a,b}$, A.J. Santiago$^a$, and 
J.W. Zhang$^b$}
\address{\it (a) Departamento de F\'\i sica Nuclear e Altas Energias, 
Instituto de F\'\i sica, Universidade do Estado do Rio de Janeiro, \\
Rio de Janeiro-RJ 20559-900, Brazil \\
(b) Department of Technical Physics, Peking University, Beijing 100871, 
China}
\date{\today}
\maketitle
\begin{abstract}Nuclear matter equations of state based on Skyrme, 
Myers-Swiatecki and Tondeur interactions are written as polynomials of 
the cubic root of density, with coefficients that are functions of the 
relative neutron excess $\delta$. In the extrapolation toward states far 
away from the standard one, it is shown that the asymmetry dependence of 
the critical point ($\rho_c, \delta_c$) depends on the model used. 
However, when the equations of state are fitted to the same standard 
state, the value of $\delta_c$ is almost the same in Skyrme and in 
Myers-Swiatecki interactions, while is much lower in Tondeur interaction. 
Furthermore, $\delta_c$ does not depend sensitively on the choice of the 
parameter $\gamma$ in Skyrme interaction.\end{abstract}

\pacs{\bf PACS numbers: 21.65.+f, 21.30.Fe}

\section{Introduction }

Nuclear matter is considered as an uncharged nucleon system distributed 
uniformly in the space, and nuclear matter equation of state is the 
energy per nucleon $e(\rho,\delta)$ of nuclear matter given as function 
of nucleon density $\rho$ and relative neutron excess $\delta$. The 
equation of state $e(\rho,\delta)$ is a fundamental quantity in theories 
of neutron stars and supernova explosions, as well as in studies of
nucleus-nucleus collisions at energies where nuclear compressibility 
comes into play\cite{Li}.

The main measured quantities which can provide information about equation 
of state (EOS) are the binding energies and other data of finite nuclei. 
As the finite nuclei are in states near the standard nuclear matter state, 
which is the ground state of nuclear matter with normal nucleon density 
$\rho_0$ and zero neutron excess, $\delta=0$, therefore, our knowledge 
about EOS can be confirmed experimentally only in a small region around
$\rho\sim\rho_0$ and $\delta\sim0$. In this region, the main quantities 
which specify the EOS are the coefficients $a_1$ (volume energy), $J$ 
(symmetry energy), $K_0$ (incompressibility), $L$ (density symmetry), and 
$K_s$ (symmetry incompressibility). Nowadays the quantities which are 
known with enough precision are $a_1$, $J$ and $K_0$, while the last two 
are still under investigation.

However, there is currently considerable interest in the very neutron 
rich nuclei and the energetic heavy-ion collisions where the nuclear 
matter state is beyond this region. As any direct information beyond this 
region is difficult to come by, extrapolation is inescapable and, in this 
case, a nuclear model is required. This model is fitted to binding 
energies and other data of finite nuclei at first, then applied to nuclear 
matter to derive the EOS. In this way, the obtained EOS can be considered 
as being fitted indirectly to a region around the standard state, but its 
prediction on states beyond this region should be regarded as an 
extrapolation. Obviously, the reliability of this extrapolation depends on 
the foundation of the model. In order to be reliable, the model should be 
based on a well-founded theory with as few adjustable parameters as 
possible, which are fitted to as many high accuracy measured data as 
possible. Basically, this is what we understand by model of effective 
nucleon-nucleon interaction, for example the Skyrme\cite{Brack}, the 
Myers-Swiatecki\cite{Myers96} and the Tondeur interactions\cite{Tondeur}. 
The EOS's derived from these effective interactions have analytical 
expressions which can be pictured and calculated easily, so they are used 
widely in the literature, even they are based on energy functional 
theories which are in the macroscopic level. Instead of effective 
nucleon-nucleon interaction, nuclear matter is studied also in more basic 
level by ``microscopic" potentials available in the market and by 
sophisticated many-body theories for many 
years\cite{Friedman}\cite{Wiringa}\cite{Pethick}\cite{Akmal}\cite{Zuo}.

The purpose of the present paper is to discuss the EOS's given by 
Skyrme\cite{Brack}, Myers-Swiatecki\cite{Myers96} and Tondeur
interactions\cite{Tondeur}, in comparison with the microscopic 
calculations. In Section 2, the nuclear EOS given by these interactions 
is presented. The equilibrium condition and the properties of standard 
nuclear matter are discussed in Sec. 3, while the predictions for nuclear 
matter away from the standard state are given in Sec. 4. Sec. 5 makes 
comparison with some microscopic calculations. In Sec. 6 a short 
discussion and summary are addressed. Appendix A presents a specific 
discussion on Myers-Swiatecki interaction and Appendix B gives some 
formulas used in Sec. 4 to calculate the interaction parameters from 
standard nuclear matter quantities.

\section{Nuclear equation of state }

The nuclear energy $E_N$ of a nucleus can be written as
\begin{equation}\label{EN} E_N=\int d^3r{\cal E}_N,\end{equation}
where the nuclear energy density functional ${\cal E}_N$ can be written 
with enough generality as
\begin{equation}\label{calEN}
 {\cal E}_N=\rho({\bf r})e(\rho,\delta)+{\cal E}_{GD},\end{equation}
where
\begin{equation}\label{calEGD} {\cal E}_{GD}=\frac{1}{2}Q_1(\nabla\rho)^2
  +Q_2\Big[(\nabla\rho_n)^2+(\nabla\rho_p)^2\Big]\end{equation}
is the gradient density dependent term. In the above equation, 
$\rho=\rho_n+\rho_p$, $\rho_n$ and $\rho_p$ are the neutron and proton 
densities respectively,
\begin{equation}\label{delta} \delta=\frac{\rho_n-\rho_p}{\rho}
\end{equation}
the relative neutron excess, $Q_1$ and $Q_2$ the model parameters related 
to the finite size and the surface effects of nuclei.

The EOS $e(\rho, \delta)$ depends on the model of interaction, while the 
functional ${\cal E}_{GD}$ depends also on the model of nuclei. Eq. 
(\ref{calEGD}) is exact for Skyrme and Tondeur interactions, and is 
approximate for Myers-Swiatecki interaction (see Appendix A). However, the 
functional ${\cal E}_{GD}$ is irrelevant to the present discussion, since 
it is irrelevant to the nuclear matter property. In the following the EOS's 
based on Skyrme, Myers-Swiatecki and Tondeur interactions will be given.

\subsection{Skyrme interaction }
The EOS based on Skyrme interaction can be written as
\begin{equation}\label{EoSSk} e^{Sk}(\rho, \delta)=
     T\Big[D_2^{Sk}(\delta)\Big(\frac\rho\rho_0\Big)^{2/3}
          -D_3^{Sk}(\delta)\Big(\frac\rho\rho_0\Big)^{3/3}
          +D_5^{Sk}(\delta)\Big(\frac\rho\rho_0\Big)^{5/3}
 +D_\gamma^{Sk}(\delta)\Big(\frac\rho\rho_0\Big)^{\gamma/3}\Big]\;,
\end{equation}
where $\rho_0=3/4\pi r_0 ^3$, $r_0$ the nuclear radius constant, $T$ an 
appropriate constant with dimension of energy such that the $D$ 
coefficients are dimensionless, and $\gamma$ a model parameter. It is 
convenient to choose $T$ as the Fermi energy of standard nuclear matter,
\begin{equation}\label{T}
 T=\frac{\hbar^2}{2m}\Big(\frac{3\pi^2}2\rho_0\Big)^{2/3},\end{equation}
where  $m$ is the nucleon mass. The $D$ coefficients are
\begin{equation}\label{D2Sk} D_2^{Sk}(\delta)=
  \frac3{10}\big[(1+\delta)^{5/3}+(1-\delta)^{5/3}\big],\end{equation}
\begin{equation}\label{D3Sk} D_3^{Sk}(\delta)=-\frac38\frac{\rho_0}T t_0
  \Big[1-\frac23\Big(x_0+\frac12\Big)\delta^2\Big],\end{equation}
\begin{equation}\label{D5Sk}
  D_5^{Sk}(\delta)=\frac3{10}\Big(\frac{3\pi^2}2\Big)^{2/3}
  \frac{\rho_0^{5/3}}T\Big\{s_1\Big[(1+\delta)^{5/3}
  +(1-\delta)^{5/3}\Big]
+\frac12s_2\Big[(1+\delta)^{8/3}+(1-\delta)^{8/3}\Big]\Big\}\;,
\end{equation}
\begin{equation}
 \label{DgSk} D_\gamma^{Sk}(\delta)=\frac1{16}\frac{\rho_0^{\gamma/3}}T
  t_3\Big[1-\frac23\Big(x_3+\frac12\Big)\delta^2\Big],\end{equation}
where
\begin{equation}s_1=\frac14\Big[t_1\Big(1+\frac{x_1}2\Big)
     +t_2\Big(1+\frac{x_2}2\Big)\Big],\,\,\,\,
 s_2=\frac14\Big[t_2\Big(x_2+\frac12\Big)
     -t_1\Big(x_1+\frac12\Big)\Big],\end{equation}
and $t_0$, $t_1$, $t_2$, $t_3$, $x_0$, $x_1$, $x_2$, $x_3$, $\gamma$ are 
the interaction parameters. It is worthwhile to note that, among these 
interaction parameters, only $t_0$, $t_3$, $x_0$, $x_3$, $s_1$, $s_2$, 
and $\gamma$ appear in the EOS and thus are relevant to the nuclear 
matter properties. Beside these interaction parameters, there is another 
interaction parameter $W_0$\cite{Brack} that appears only in the 
coefficients $Q_1$ and $Q_2$ and thus is irrelevant to the EOS.

\subsection{Myers-Swiatecki interaction }
The EOS based on Myers-Swiatecki interaction can be written 
as\cite{WCS}\cite{Myers98}
\begin{equation}\label{EoSMS} e^{MS}(\rho, \delta)=
     T\Big[D_2^{MS}(\delta)\Big(\frac\rho\rho_0\Big)^{2/3}
          -D_3^{MS}(\delta)\Big(\frac\rho\rho_0\Big)^{3/3}
          +D_5^{MS}(\delta)\Big(\frac\rho\rho_0\Big)^{5/3}\Big],
\end{equation}
where
\begin{eqnarray}\label{D2MS} D_2^{MS}(\delta)&=&\frac3{10}(1-\gamma_l )
               \big[(1+\delta)^{5/3}+(1-\delta)^{5/3}\big]\nonumber\\
                & &-\frac3{20}\gamma_u \times\left\{
   \begin{array}{ll}
    5(1+\delta)^{2/3}(1-\delta)-(1-\delta)^{5/3},\qquad
                            & \mbox{for}\,\, \delta\geq 0, \\
    5(1+\delta)(1-\delta)^{2/3}-(1+\delta)^{5/3},\qquad
                            & \mbox{for}\,\, \delta\leq 0,
   \end{array}\right.\end{eqnarray}
\begin{equation}\label{D3MS}
 D_3^{MS}(\delta)=\frac12\alpha(1-\xi\delta^2),\end{equation}
\begin{equation}\label{D5MS} D_5^{MS}(\delta)=
 \frac3{10}\Big\{B_l\Big[(1+\delta)^{8/3}+(1-\delta)^{8/3}\Big]
 +B_u(1-\delta^2)\Big[(1+\delta)^{2/3}+(1-\delta)^{2/3}\Big]\Big\}.
\end{equation}
In the above equations, $\alpha$, $B_l$, $B_u$, $\gamma_l$, $\gamma_u$, 
and $\xi$ are the interaction parameters. In addition to these parameters, 
there is another one $a$\cite{Myers96}, the Yukawa range of force, that is 
irrelevant to the EOS, as it appears only in the coefficients $Q_1$ and 
$Q_2$.

\subsection{Tondeur interaction }
The EOS based on Tondeur interaction can be written as
\begin{equation}\label{EoSTo} e^{To}(\rho, \delta)=
    T\Big[D_2^{To}(\delta)\Big(\frac\rho\rho_0\Big)^{2/3}
         -D_3^{To}(\delta)\Big(\frac\rho\rho_0\Big)^{3/3}
     +D_\gamma^{To}(\delta)\Big(\frac\rho\rho_0\Big)^{\gamma/3}\Big],
\end{equation}
where
\begin{equation}\label{D2To} D_2^{To}(\delta)=
  \frac3{10}\big[(1+\delta)^{5/3}+(1-\delta)^{5/3}\big]
  +\frac{\rho_0^{2/3}c}T\,\delta^2,\end{equation}
\begin{equation}\label{D3To} D_3^{To}(\delta)=-\frac{\rho_0 a}T,
\end{equation}
\begin{equation}\label{DgTo}
 D_\gamma^{To}(\delta)=\frac{\rho_0^{\gamma/3}b}T.\end{equation}
In the above equations, $a$, $b$, $c$, and $\gamma$ are the interaction
parameters. In addition, there are another two interaction parameters $d$ 
and $\eta$\cite{Tondeur}, that are irrelevant to the present discussion
 as they appear only in $Q_1$ and $Q_2$.

\section{Standard nuclear matter }

The EOS given in the last section can be written generally as
\begin{equation}\label{EoS}
 e(\rho, \delta)=T\Big[D_2(\delta)\Big(\frac\rho\rho_0\Big)^{2/3}
                -D_3(\delta)\Big(\frac\rho\rho_0\Big)^{3/3}
                +D_5(\delta)\Big(\frac\rho\rho_0\Big)^{5/3}
       +D_\gamma(\delta)\Big(\frac\rho\rho_0\Big)^{\gamma/3}\Big].
\end{equation}
The equilibrium condition $\partial e/\partial\rho|_0=0$, by which the 
standard state $\rho=\rho_0$ at $\delta=0$ is defined, gives the following 
relationship among $D_2(0)$, $D_3(0)$, $D_5(0)$, $D_\gamma(0)$ and $\gamma$:
\begin{equation}\label{stable0}
 2D_2(0)-3D_3(0)+5D_5(0)+\gamma D_\gamma(0)=0.\end{equation}
The $5$ quantities of nuclear matter $a_1$, $K_0$, $J$, $L$, and $K_s$ 
can be expressed as:
\begin{equation}\label{a1} a_1=-e(\rho_0, 0)=-\frac T3\big[D_{20}
-2D_{50}-(\gamma-3)D_{\gamma 0}\big],\end{equation}
\begin{equation}\label{K0}
 K_0=9\rho_0^2\frac{\partial^2e}{\partial\rho^2}\Big |_0
    =T\big[-2D_{20}+10D_{50}+\gamma(\gamma-3)D_{\gamma 0}\big],
\end{equation}
\begin{equation}\label{J}
 J=\frac12\frac{\partial^2e}{\partial\delta^2}\Big |_0
  =T\big[D_{22}-D_{32}+D_{52}+D_{\gamma 2}\big],\end{equation}
\begin{equation}\label{L} L=\frac32\rho_0
   \frac{\partial^3e}{\partial\rho\partial\delta^2}\Big |_0
  =T\big[2D_{22}-3D_{32}+5D_{52}+\gamma D_{\gamma 2}\big],\end{equation}
\begin{equation}\label{Ks}
 K_s=\frac92\rho_0^2
     \frac{\partial^4e}{\partial\rho^2\partial\delta^2}\Big |_0
    =T\big[-2D_{22}+10D_{52}+\gamma(\gamma-3)D_{\gamma 2}\big],
\end{equation}
where
\begin{equation}D_{i0}=D_i(0),\,\,\,\,
 D_{i2}=\frac 12\frac{\partial^2D_i}{\partial\delta^2}|_0,
 \,\,\,\, i=2, 3, 5, \gamma.\end{equation}
Relation (\ref{stable0}) is used in obtaining Eqs. (\ref{a1}) and 
(\ref{K0}), from which the following formulas can be derived:
\begin{equation}\label{K0a1a}K_0=15a_1+[3D_{20}+(\gamma-5)(\gamma-3)
 D_{\gamma0}]T,\end{equation}
\begin{equation}\label{K0a1b}
K_0=3\gamma a_1+[(\gamma-2)D_{20}-2(\gamma-5)D_{50}]T.\end{equation}

The specific discussion for Skyrme, Myers-Swiatecki and Tondeur 
interactions will be given in the following.

\subsection{Skyrme interaction }
For Skyrme interaction, the following relationship can be obtained from 
Eq. (\ref{stable0}):
\begin{equation}\label{stableSk}\frac98\frac{\rho_0t_0}T+\frac\gamma{16}
 \frac{\rho_0^{\gamma/3}t_3}T+3\Big(\frac{3\pi^2}2\Big)^{2/3}
 \frac{\rho_0^{5/3}}T\big(s_1+\frac12s_2\big)+\frac65=0.\end{equation}
Therefore, among $7$ parameters $t_0$, $t_3$, $x_0$, $x_3$, $s_1$,
$s_2$, and $\gamma$, only $6$ of them are free. Considering this
relation, it can be shown that $a_1$, $K_0$, $J$, $L$ and $K_s$ are 
independent each other, in the Skyrme EOS.

A relation connecting $t_3$ to $a_1$ and $K_0$ can be obtained from
Eq. (\ref{K0a1a}),
\begin{equation}\label{K0Sk}K_0=15a_1+\frac95T
 +\frac{(\gamma-5)(\gamma-3)}{16}\rho_0^{\gamma/3}t_3.\end{equation}
If $t_3=0$, as $a_1\sim16MeV$ and $T\sim37MeV$ are well-known from
measurements, this formula gives the estimation $K_0\sim 306 MeV$. 
Therefore, in order to have $K_0$ value lower than $306 MeV$, the fourth 
term $(\rho/\rho_0)^{\gamma/3}$ in the Skyrme EOS is needed.

\subsection{Myers-Swiatecki interaction }
For Myers-Swiatecki interaction, $D^{MS}_\gamma(\delta)=0$, all the 
$\gamma$-dependent terms in Eqs. (\ref{stable0})-(\ref{Ks}) do not appear. 
The equilibrium condition (\ref{stable0}) can be transformed into the 
following relation among $\alpha$, $B$ and $\overline\gamma$:
\begin{equation}\label{aBgamma}
 5\alpha-10B-4(1-\overline\gamma)=0,\end{equation}
where $B$ and $\overline\gamma$ are defined respectively as
\begin{equation}B_{l,u}=\frac12(1\mp\zeta)B,\,\,\,\,
 \gamma_{l,u}=\frac12(1\mp\zeta)\overline\gamma.\end{equation}
Therefore, there are only $4$ independent interaction parameters in the 
Myers-Swiatecki EOS, $\alpha$, $B$, $\xi$ and $\zeta$, if $\overline\gamma$ 
is solved from Eq. (\ref{aBgamma}) as a function of $\alpha$ and $B$. 
Correspondingly, there are only $4$ independent variables among $a_1$, 
$K_0$, $J$, $L$ and $K_s$ in the Myers-Swiatecki EOS. Actually, the 
following relationship can be derived:
\begin{equation}\label{KsMS} \frac{K_s}T=\frac{4B(1
 +\overline\gamma)}{4B+\overline\gamma}\Big[1-\frac{10B
 +\overline\gamma}{2B(1+\overline\gamma)}\frac{3J-L}T\Big],\end{equation}
where
\begin{equation}B=\frac5{18}\frac{K_0-6a_1}T,\end{equation}
\begin{equation}\overline\gamma=1-\frac59\frac{K_0-15a_1}T.\end{equation}

Furthermore, formula (\ref{K0a1a}) for the Myers-Swiatecki EOS becomes
\begin{equation}K_0=15a_1+\frac95(1-\overline\gamma)T.\end{equation}
For $\overline\gamma=0$, Myers-Swiatecki interaction is reduced to
Seyler-Blanchard interaction\cite{Seyler} and this formula gives the
estimation $K_0\sim 306 MeV$, the same as that discussed for Skyrme 
interaction. Hence, $\overline\gamma$-dependent terms in 
Myers-Swiatecki EOS are required, in order to obtain $K_0$ lower than 
$306 MeV$\cite{WCS}.

\subsection{Tondeur interaction }
For Tondeur interaction, $D_5(\delta)=0$, the term involving $D_5(0)$ 
in Eq. (\ref{stable0}) as well as all the terms involving $D_{50}$ and 
$D_{52}$ in Eqs. (\ref{a1})-(\ref{Ks}) do not appear. The equilibrium 
condition (\ref{stable0}) now is a relation among $a$, $b$ and $\gamma$:
\begin{equation}\frac{3\rho_0a}T+\frac{\gamma\rho_0^{\gamma/3}b}T
+\frac65=0.\end{equation}
In this case, there are only $3$ independent interaction parameters in
Tondeur EOS, for example $a$, $c$ and $\gamma$. Correspondingly, there 
are only $3$ free variables in $a_1$, $K_0$, $J$, $L$ and $K_s$, for 
example $a_1$, $K_0$ and $J$, since it can be shown that
\begin{equation}\label{LKsTo}L=2J,\,\,\,\, K_s=-2J.\end{equation}

In addition, the following relationship among $K_0$, $a_1$ and $\gamma$ 
can be written for Tondeur EOS from Eq. (\ref{K0a1b}):
\begin{equation}\label{K0To}
 K_0=3\gamma a_1+\frac35(\gamma-2)T.\end{equation}
From $a_1\sim16MeV$, $T\sim37MeV$ and $K_0\sim220MeV$, it can be 
evaluated that the appropriate integer is $\gamma=4$, as given by 
Tondeur\cite{Tondeur}. In this case, {\it i.e.}, if $\gamma=4$ is chosen, 
there are only two interaction parameters to be freely adjusted in the 
data fit, for example $a$ and $c$. Correspondingly, there are only two 
independent variables in $a_1$, $K_0$, $J$, $L$ and $K_s$, for example 
$a_1$ and $J$, when $K_0$ is calculated by Eq. (\ref{K0To}). From 
$a_1\sim16MeV$, $T\sim37MeV$ and $\gamma=4$ we can evaluate 
$K_0\sim236MeV$. It is worthwhile to note that the value given by Tondeur 
is $K_0=235.8MeV$\cite{Tondeur}.

\vskip 0.5cm 
The equilibrium condition is checked by calculating the expression on the 
lefthand side of Eq. (\ref{stable0}) for Skyrme, Myers-Swiatecki and 
Tondeur interactions, using the interaction parameters and the nuclear 
radius constant $r_0$ given in Refs. \cite{Brack}, \cite{Myers96}, and 
\cite{Tondeur}, respectively. These parameters will be referred to as the 
original interaction parameters thereafter. Besides, the following 
physical constants\cite{Moller} are used in the present calculation: 
$\hbar c=197.32891MeV\cdot fm$, $m=938.90595MeV/c^2$.

The calculated values are given as $EC$ in the second column of Table 1. 
It shows that the equilibrium condition of standard nuclear matter is 
fulfilled in the data fit to determine the original parameters of Skyrme 
(1st-5th row), Myers-Swiatecki (6th row), and Tondeur interactions 
(7th row), respectively.

The standard nuclear matter properties $a_1$, $K_0$, $J$, $L$, and $K_s$, 
calculated from Skyrme (1st-5th row), Myers-Swiatecki (6th row), and 
Tondeur interactions (7th row) respectively, are also given in the 5th-9th 
column of Table 1, all in $MeV$. In this table, $\gamma$ is a model 
parameter in Eqs. (\ref{EoSSk}) and (\ref{EoSTo}) for Skyrme and Tondeur 
interactions respectively, $r_0$ the nuclear radius constant used in the 
respective interaction, in $fm$.

As a comparison, the last two rows of Table 1 (labeled by CWS) present the 
result obtained by fitting $a_1$, $K_0$, $J$, $L$, and $K_s$ directly to 
nuclear masses\cite{CWS}. It can be seen that these quantities have values 
close each other, except the case SIII, where the value of $K_0$ and $K_s$ 
is far away from others. The average over the 2nd to 7th row gives 
$a_1=15.97MeV$, $K_0=234.4MeV$, $J=29.25MeV$, $L=48.63MeV$, and 
$K_s=-126.9MeV$.

\section{Nuclear matter away from the standard state }

The nuclear matter state with zero pressure and minimum energy per nucleon 
can be solved from the following equation:
\begin{equation}\label{stable}
 \frac{\partial e}{\partial\rho}=0.\end{equation}
Usually there are several solutions, we should choose that one has minimum 
energy per nucleon. This solution gives density as function of $\delta$:
\begin{equation}\label{rhom}\rho_m=\rho_m(\delta).\end{equation}
For $\delta=0$, Eq. (\ref{stable}) is reduced to the equilibrium
condition of standard nuclear matter, we have
\begin{equation}\rho_m(0)=\rho_0.\end{equation}

The incompressibility of non-equilibrium nuclear matter, which is of 
interest in many applications, can be defined as\cite{Myers98}
\begin{equation}\label{K}
 K(\rho,\delta)=9\frac{\partial P}{\partial\rho},\end{equation}
where $P=\rho^2\partial e/\partial\rho$ is the pressure. Along the line of 
minimum (\ref{rhom}), this $K(\rho,\delta)$ becomes
\begin{equation}K_m(\delta)=9\;\big[\rho^2
 \frac{\partial^2e}{\partial\rho^2}\big]_{\rho=\rho_m}.\end{equation}
At the standard state $(\rho_0,0)$ we have $K_m(0)=K_0$. At the critical 
point $(\rho_c,\delta_c)$, where the maximum and the minimum are 
coincident, the curvature of $e(\rho,\delta_c)$ versus $\rho$ changes sign 
and $K_m(\delta_c)=0$. So $K_m(\delta)$ starts with $K_0$ and ends at $0$ 
when $\delta$ increases along the line of minimum. In addition, the 
generalized symmetry energy of non-equilibrium nuclear matter can be 
defined as\cite{Wiringa}\cite{Li}\cite{Zuo}
\begin{equation}\label{Jrho}J(\rho)=\frac12
 \frac{\partial^2e}{\partial\delta^2}\Big |_{\delta=0}.\end{equation}
In term of this quantity, the usual symmetry energy $J$ can be expressed as
\begin{equation}J=J(\rho_0).\end{equation}

For nuclear matter not far away from the standard state $(\rho_0,0)$, the 
EOS can be written approximately as\cite{CWS}
\begin{equation}\label{EoSa}
 e(\rho,\delta)\approx -a_1+\frac 1{18}\big(K_0+K_s\delta^2\big)
 \Big(\frac{\rho-\rho_0}{\rho_0}\Big)^2+\Big[J+\frac L3
 \Big(\frac{\rho-\rho_0}{\rho_0}\Big)\Big]\delta^2.\end{equation}
In this approximation, we have
\begin{equation}\label{Kapprox}K(\rho,\delta)
 \approx(K_0+K_s\delta^2)\Big(\frac\rho{\rho_0}\Big)^2,\end{equation}
\begin{equation}\label{Jrhoa}J(\rho)\approx J+\frac L3\frac{\rho
 -\rho_0}{\rho_0}+\frac{K_s}{18}\Big(\frac{\rho-\rho_0}{\rho_0}\Big)^2.
\end{equation}

Using Eq.(\ref{EoSa}), the following solutions can be obtained:
\begin{equation}\label{rhoma}\rho_m\approx\rho_0\Big(1-
 \frac{3L}{K_0}\delta^2\Big),\end{equation}
\begin{equation}\label{ema}e_m\approx -a_1+J\delta^2,\end{equation}
\begin{equation}\label{Kma}K_m\approx K_0+K_s\delta^2,\end{equation}
where only the linear term in $\delta^2$ is kept. The systematics of 
nuclear central densities\cite{WCS99} based on elastic electron scattering 
data\cite{deVries}\cite{Fricke} and muonic atom spectroscopy 
data\cite{Fricke} provide a direct evidence for Eq. (\ref{rhoma}).

Thus, in the plot $e(\rho,\delta)$ versus $\rho$, we have the geometric 
meaning of $a_1$, $K_0$, $J$, $L$, and $K_s$: the standard state is at the 
minimum point $\rho_m=\rho_0$ with depth $a_1$ and curvature proportional 
to $K_0$; when the minimum is moved with increasing $\delta$ from $0$, the 
decrease of $\rho_m$ is controlled by $3L/K_0$, the increase of depth is 
controlled by $J$, while the decrease of curvature is controlled by $-K_s$. 
Therefore, the quantities $a_1$, $K_0$, $J$, $L$, and $K_s$ are 
characteristics of nuclear matter not only at standard state but also at 
the state not far away from the standard one. In this way, the interaction 
with different value of these quantities will predict different properties 
of nuclear matter that are not far away from the standard state.

The exact solution $\rho_m(\delta)$ depends on the interaction. The 
analytic solution is possible for SIII, Ska, Myers-Swiatecki, and Tondeur 
interactions, while the numerical solution is appropriate for SkM, SkM*, 
and RAPTP interactions.

For Myers-Swiatecki interaction, (\ref{stable}) is a cubic equation which 
gives
\begin{equation} \Big(\frac{\rho_m}{\rho_0}\Big)^{1/3}
 =2s_0sin\Big(\frac\pi6+\frac\theta3\Big),\end{equation}
where
\begin{equation}s_0=\Big[\frac15\frac{D_3(\delta)}{D_5(\delta)}
 \Big]^{1/2},\,\,\,\,cos\theta=\frac1{5s_0^3}
 \frac{D_2(\delta)}{D_5(\delta)}.\end{equation}
The superscript $MS$ for Myers-Swiatecki's $D$ is dropped for simplicity. 
The critical $\delta_c$, where the maximum and the minimum is coincident, 
is determined by
\begin{equation}D_2(\delta_c)=5s_0^3D_5(\delta_c),\end{equation}
which corresponds to $\theta=0$ and
\begin{equation}\rho_c=\rho_m(\delta_c)=s_0^3\rho_0.\end{equation}

For Tondeur interaction with $\gamma=4$, (\ref{stable}) is a quadratic 
equation which gives
\begin{equation} \Big(\frac{\rho_m}{\rho_0}\Big)^{1/3}=\frac1{8D_4}
 \big\{3D_3+[9D_3^2-32D_4D_2(\delta)]^{1/2}\big\},\end{equation}
where $D_3$ and $D_4$ are numbers, the superscript $To$ for Tondeur's $D$ 
is dropped also. The critical point is given by
\begin{equation}\label{CritPTo} 9D_3^2-32D_4D_2(\delta_c)=0,\,\,\,\,
 \rho_c=\Big(\frac{3D_3}{8D_4}\Big)^3\rho_0=-\Big(\frac{3a}{8b}\Big)^3.
\end{equation}

As the location $\rho_0$ and depth $a_1$ are different for different 
equation of state, as shown in Table 1, a way to make comparison is to 
plot the normalized energy per nucleon $e/a_1$ as a function of the 
relative nucleon density $\rho/\rho_0$ for given $\delta$. Fig. 1a shows 
this $e/a_1$ versus $\rho/\rho_0$ for $\delta=0$, calculated by the 
various Skyrme interactions (solid lines), Myers-Swiatecki interaction 
(dot-dashed line), and Tondeur interaction (dashed line). The solid lines 
in the righthand side of the plot, from top to bottom, correspond to SIII, 
Ska, RATP, SkM, and SkM* interactions respectively. The difference between 
SkM and SkM* is negligible and Myers-Swiatecki is almost coincident with 
RATP. This sequence is just the decreasing sequence of $K_0$'s value, as 
shown in Table 1: the smaller value of $K_0$, the smaller curvature of the 
curve at the standard state, thus the softer the EOS.

A natural question is: what is the difference among these EOS's, if the 
standard state is the same with same location $\rho_0$, depth $a_1$, 
curvature $\sim K_0$ and so on? In order to make this comparison, the 
interaction parameters should be readjusted according to chosen $\rho_0$, 
$a_1$, $K_0$ and so on. For the value of $\rho_0$, we choose $r_0=1.140fm$ 
which is well determined by the data fit to nuclear charge 
radii\cite{Buchinger} extracted from elastic electron scattering
data\cite{deVries}. In addition, we can choose the value of $a_1$, $K_0$, 
$J$, $L$, and $K_s$ in an appropriate way. In this case, Eq. (\ref{KsMS}) 
should be fulfilled for Myers-Swiatecki interaction, while Eqs. 
(\ref{LKsTo}) and (\ref{K0To}) should be fulfilled for Tondeur interaction. 
The chosen values used to calculate the interaction parameters are listed 
in Table 2, while the formulas used to perform this calculation are given 
in Appendix B. Among these values, $a_1$ and $J$ are the average values 
given in the last section, $K_0$ and $L$ are calculated by Eqs. 
(\ref{K0To}) and (\ref{LKsTo}). $K_s$ is calculated by Eq. (\ref{LKsTo}) 
in Tondeur's case while by Eq. (\ref{KsMS}) in Myers-Swiatecki's case. In 
Skyrme's case, $K_s$ can be chosen from either Myers-Swiatecki's or 
Tondeur's value, there is no significant difference in the calculated 
result which will be shown in the following.

The calculated Skyrme interaction parameters are given in Table 3. The 
Myers-Swiatecki interaction parameters are calculated as:
\begin{equation}\label{paraMS}\alpha=2.06285,\;\; B=1.05232,\;\;
 \overline\gamma=1.05222,\;\;\xi=0.12333,\;\; \zeta=0.37363.
\end{equation}
For $\gamma=4$, Tondeur interaction parameters are calculated as:
\begin{equation}\label{paraTo} a=-672.13MeVfm^3,\,\,
 b=799.71MeVfm^4,\,\,c=99.116MeVfm^2.\end{equation}
These parameters will be referred to as the readjusted interaction 
parameters thereafter.

As a comparison with the result calculated by original interaction 
parameters, Fig. 1b plots $e$ versus $\rho/\rho_0$ for $\delta=0$, 
calculated by readjusted interaction parameters. It can be seen that now 
there is almost no difference among these EOS's for $0.4<\rho/\rho_0<1.6$.

Fig. 2 displays in (a) the normalized nuclear incompressibility $K/K_0$ 
versus the relative nucleon density $\rho/\rho_0$ calculated by original 
interaction parameters at $\delta=0$, and in (b) $K$ versus $\rho/\rho_0$ 
calculated by readjusted interaction parameters at $\delta=0$ of the 
various Skyrme interactions (solid lines), Myers-Swiatecki interaction 
(dot-dashed line), and Tondeur interaction (dashed line). Considering the 
righthand side of the plot, in (a), the solid lines, from top to bottom, 
correspond to SIII, Ska, RATP, SkM, and SkM* interactions respectively. 
The difference between SkM and SkM* is negligible. In (b), the solid 
lines, from top to bottom, are due to SkM, SkM*, RATP, Ska, and SIII, 
where SkM* is identical to SkM. Tondeur is coincident with Ska; SkM and 
RATP almost overlap. It can be seen from this figure that the difference 
among these curves is negligible for $\rho/\rho_0<1.2$. This is expected 
from Eq. (\ref{Kapprox}) which shows that the curve is determined 
essentially by $K_0$.

Fig. 3 depicts in (a) the normalized symmetry energy $J(\rho)/J$ versus 
the relative nucleon density $\rho/\rho_0$ calculated by original 
interaction parameters, in (b) $J(\rho)$ versus $\rho/\rho_0$ calculated 
by readjusted interaction parameters of the various Skyrme interactions 
(solid lines), Myers-Swiatecki interaction (dot-dashed line), and Tondeur 
interaction (dashed line). Considering the righthand side of the plot, in 
(a), the solid lines, from top to bottom, correspond to Ska, SkM, SkM*, 
RATP, and SIII interactions respectively. The difference between SkM and 
SkM* for $\rho/\rho_0<1.3$ is negligible. In addition, the 
Myers-swiatecki's is almost coincident with that of SkM's. In (b), the 
difference among these curves is negligible. This is expected from 
Eq.(\ref{Jrhoa}), which shows that, for the density $\rho$ is not far 
away from $\rho_0$, the symmetry energy $J(\rho)$ is determined 
essentially by $J$, $L$ and $K_s$, and these quantities ($J$ and $L$) are 
the same or almost the same ($K_s$) for the readjusted interaction 
parameters. 

Using the interaction parameters, we can calculate $\rho_m(\delta)$, 
$e_m(\delta)$, and $K_m(\delta)$ along the equilibrium line. The result 
is shown in Figs. 4-6, while the critical point value 
$(\rho_c,\delta_c)$ is listed in Table 4, for Skyrme, Myers-Swiatecki, 
and Tondeur interactions respectively.

Fig. 4 shows the equilibrium density $\rho_m$ as function of the relative 
neutron excess $\delta$, calculated by (a) original parameters and (b) 
readjusted parameters of the various Skyrme interactions (solid lines), 
Myers-Swiatecki interaction (dot-dashed line), and Tondeur interaction 
(dashed line). The solid lines, in the middle range of $\delta$, from top 
to bottom, in (a), correspond to SIII, RATP, SkM, SkM*, and Ska 
interactions respectively, the difference between SkM and SkM* is very 
small. In (b), the solid lines from top to bottom correspond to Ska, RATP, 
SkM, SkM*, and SIII interactions respectively. SkM and SkM* are the same 
whereas RATP, SkM and Myers-Swiatecki almost overlap.

Fig. 5 gives the equilibrium energy per nucleon $e_m$ as function of the 
relative neutron excess $\delta$, calculated by (a) original parameters 
and (b) readjusted parameters of the various Skyrme interactions (solid 
lines), Myers-Swiatecki interaction (dot-dashed line), and Tondeur 
interaction (dashed line). The solid lines, in the righthand side, in 
(a), correspond to RATP, SkM, SkM*, SIII, and Ska interactions, from top 
to bottom, respectively. SIII, Ska and Tondeur almost overlap, whereas 
the difference between SkM and SkM* is negligible. In (b), the solid 
lines from top to bottom correspond to SIII, SkM, SkM*, RATP, and Ska 
interactions, respectively, where SkM and SkM* are the same. SkM, RATP 
and Ska are almost coincident.

Fig. 6 plots the equilibrium incompressibility $K_m$ as function of the 
relative neutron excess $\delta$, calculated by (a) original parameters 
and (b) readjusted parameters of the various Skyrme interactions (solid 
lines), Myers-Swiatecki interaction (dot-dashed line), and Tondeur 
interaction (dashed line). The solid lines in (a) correspond to SIII, 
RATP, Ska, SkM*, and SkM interactions, from top to bottom in the middle 
range of $\delta$, respectively. Ska, SkM* and SkM almost overlap. In (b), 
the solid lines, from top to bottom, correspond to Ska, RATP, SkM, SkM*, 
and SIII interactions respectively; SkM and SkM* are the same.

\section{Comparison with microscopic calculations}

In order to provide additional elements about the confidence on the 
effective interactions discussed above, it is interesting to make a 
comparison with some microscopic calculations which are based on a more 
fundamental level of theories as well as on very different physical input. 
In Fig. 7 the present predictions for the pure neutron matter EOS are 
compared with the theoretical estimates of Friedman and 
Pandharipande\cite{Friedman}, obtained from a variational framework based 
on the Urbana $v_{14}$ two-nucleon potential plus three-nucleon 
interaction model of Lagaris and Pandharipande\cite{Lagaris}. The neutron 
matter EOS $e(\rho,1)$ versus nucleon density $\rho$ is calculated by (a) 
original parameters and (b) readjusted parameters of the various Skyrme 
interactions (solid lines), Myers-Swiatecki interaction (dot-dashed 
line), and Tondeur interaction (dashed line). Black diamonds denote the 
data taken from Ref. \cite{Friedman}. The solid lines, in the right hand 
side of (a), correspond to Ska, SkM, SkM*, RATP, and SIII from top to 
bottom. In (b), the solid lines, from top to bottom, correspond to SIII, 
SkM, SkM*, RATP, and Ska, where SkM and SkM* are identical. It seems 
that in the low density region the neutron matter EOS's calculated by 
original parameters are closer to microscopic results than those by 
readjusted parameters. However, the situation is different if the density 
is extended to higher region (see Fig. 9).

It is worthwhile to note that the generalized Skyrme interaction FPS21 
proposed by Pethick, Ravenhall and Lorenz\cite{Pethick} has the property 
that it is a good fit to both the nuclear and neutron matter calculations 
of Friedman and Pandharipande. In this sense, Fig. 7 may be regarded also 
as a comparison beetween our results and those of FPS21.

Even the EOS's based on effective interactions and energy functional 
theories discussed in the present work are essentially nonrelativistic, 
it is still interesting to see how they behave in the high density 
region. Fig. 8 gives the symmetric nuclear matter EOS's $e(\rho,0)$ up to 
about $10\rho_0$, calculated by (a) original parameters and (b) 
readjusted parameters of the various Skyrme interactions (solid lines), 
Myers-Swiatecki interaction (dot-dashed line), and Tondeur interaction 
(dashed line). The full dots stand for the results taken from Ref. 
\cite{Wiringa}, which is a microscopic calculation of EOS for dense 
nuclear and neutron matter based on the Argonne $v_{14}$ two-nucleon 
potential plus Urbana VII three-nucleon potential. The crosses denote the 
results taken from Ref. \cite{Akmal}, which studied the properties of 
dense nucleon matter and the structure of neutron stars, using 
variational chain summation methods and the new Argonne $v_{18}$ 
two-nucleon interaction and the Urbana model IX of three-nucleon 
interaction as well as the relativistic boost correction to the 
two-nucleon interaction. The solid lines, in the right hand side of (a), 
correspond to SIII, Ska, RATP, SkM, and SkM*, from top to bottom, where 
SkM and SkM* are identical. In (b), the solid lines, from top to bottom, 
correspond to SkM, SkM*, RATP, Ska, and SIII, where SkM and SkM* are 
identical. The Tondeur's is very close to the Ska's.

Fig. 9 is the same as Fig. 8 but for pure neutron matter EOS $e(\rho,1)$. 
The solid lines, in the right hand side of (a), correspond to Ska, RATP, 
SkM, SkM*, and SIII, from top to bottom. In (b), the solid lines, from 
top to bottom, correspond to SIII, SkM, SkM*, RATP, and Ska, where SkM 
and SkM* are identical.

\section{Discussion and Summary }

A discussion of the nuclear matter EOS's based on Skyrme, Myers-Swiatecki 
and Tondeur interactions is given in this paper. The equations are in the 
form of polynomials in the cubic root of density, with coefficients that
are functions of the relative neutron excess and depend on the model of 
interaction. 

Most of the discussion about the nuclear EOS, up to now, focus at states 
around standard state, {\it i.e.} about the quantities $a_1$, $J$, $L$, 
$K_0$, and $K_s$; especially $K_0$ in supernova explosion and neutron 
star calculations and $K_s$ in heavy ion collisions. However, even these 
quantities or equivalently the interaction parameters were 
well-determined by the measured data of nuclei, mainly the nuclear 
masses, the extrapolation to states far away from standard state is 
still an open problem. It is seen that the difference among these EOS's 
is not significant in most of the relative neutron excess range which 
is of interest for both heavy ion collisions and supernova explosion 
calculations. However, if the equations are fitted to the same standard 
state, the equation based on Tondeur interaction is softer than others 
provided the relative neutron excess is not close to $0$ \cite{CWS00}. 

 The numerical result given in Section IV shows that the asymmetry 
dependence of the critical point depends on the model used in the 
extrapolation. When the EOS is fitted to same standard state, the 
Skyrme's and the Myers-Swiatecki's $\delta_c$ are close each other, 
especially $\delta_c$ does not depend sensitively on the choice of 
$\gamma$ in Skyrme interaction. On the other hand, the Tondeur's 
$\delta_c$ is much smaller than others. This is because the value of 
the Tondeur's $\delta_c$ depends sensitively on the interaction 
parameters, as it can be seen and checked numerically from the first 
equation of (\ref{CritPTo}). In this content, in order to make a choice 
among these interactions for the extrapolation, experiments which can 
provide direct or even indirect information about nuclear matter with 
large asymmetry $\delta$ and low density $\rho$ are required.

\acknowledgements{The authors acknowledge the support from the 
Funda\c c\~ao de Amparo \`a Pesquisa do Estado do Rio de Janeiro 
(FAPERJ). C.S.W. and J.W.Z. also acknowledge the partial support by the 
National Natural Science Foundation of China and the Special Fund of 
China for the Universities with Ph.D. Programs.}

\appendix
\section{}

It will be shown here that the energy density functional ${\cal E}_{GD}$ of 
Myers-Swiatecki interaction can be written approximately in the form of Eq. 
(\ref{calEGD}). In the Thomas-Fermi model of nuclei and up to the second 
order of the localized approximation given in Ref. \cite{WCS}, we have
\begin{equation}{\cal E}_{GD}^{MS}=aI_1(r/a)F^{(1)}(r)
 +\frac{a^2}2I_2(r/a)F^{(2)}(r),\end{equation}
where $a$ is the Yukawa range of force,
\begin{equation}I_1(x)=\frac2x(1-e^{-x}),\,\,\,\,I_2(x)=2(1+2e^{-x}),
\end{equation}
\begin{equation}\label{F1}F^{(1)}(r)=T\,\Big[\epsilon_{1n}
 \frac{d\rho_n}{dr}+\epsilon_{1p}\frac{d\rho_p}{dr}\Big],\end{equation}
\begin{equation}\label{F2}F^{(2)}(r)=T\,\Big[\epsilon_{1n}
 \frac{d^2 \rho_n}{dr^2}+\epsilon_{1p}\frac{d^2 \rho_p}{dr^2}
 +\frac{2\epsilon_{2n}}{\rho _0}\Big(\frac{d\rho_n}{dr}\Big)^2
 +\frac{2\epsilon_{2p}}{\rho _0}\Big(\frac{d\rho_p}{dr}\Big)^2\Big].
\end{equation}
In the above two equations, $\epsilon_{1n}, \epsilon_{1p}, \epsilon_{2n}$ 
and $\epsilon_{2p}$ are the functionals of nucleon densities $\rho_n(r)$ 
and $\rho_p(r)$ whose specific expressions are given in Ref. \cite{WCS}. 
Using the approximation of $I_1(x)\approx 2/x$ and $I_2(x)\approx 2$ which 
are explained and employed in Ref. \cite{WCS}, the following result can be 
obtained:
\begin{equation}{\cal E}_{GD}^{MS}=a^2T\big(\epsilon_{1n}\nabla^2\rho_n
 +\epsilon_{1p}\nabla^2\rho_p\big).\end{equation}
In the simplified Myers-Swiatecki interaction, we have\cite{WCS}
\begin{equation}\epsilon_{1n}=-\Big[\alpha_l\frac{\rho_n}{\rho_0}
               +\alpha_u\frac{\rho_p}{\rho_0}\Big],\,\,\,\,
 \epsilon_{1p}=-\Big[\alpha_l\frac{\rho_p}{\rho_0}
               +\alpha_u\frac{\rho_n}{\rho_0}\Big],\,\,\,\,
 \epsilon_{2n}=\epsilon_{2p}=0,\end{equation}
where $\alpha_{l,u}=\frac12(1\pm\xi)\alpha$, thus the functional
${\cal E}_{GD}^{MS}$ can be reduced to
\begin{equation}{\cal E}_{GD}^{MS}=\frac{a^2T}{\rho_0}\big\{
 \alpha_u(\nabla\rho)^2+(\alpha_l-\alpha_u)[(\nabla\rho_n)^2
 +(\nabla\rho_p)^2]\big\}.\end{equation}
For the symmetric case with $\rho_n=\rho_p=\rho/2$, we have finally
\begin{equation}{\cal E}_{GD}^{MS}=\frac{a^2T}{2\rho_0}
 \alpha(\nabla\rho)^2.\end{equation}

\section{}

The formulas to calculate the interaction parameters from the nuclear 
matter quantities $a_1$, $K_0$, $J$, $L$, and $K_s$ will be given here for 
Skyrme, Myers-Swiatecki, and Tondeur interactions respectively. In Skyrme 
interaction, $s_1$, $s_2$, $t_3$, and $x_3$ can be calculated by the 
following equations:
\begin{equation}s_1+\frac12s_2=\Big(\frac2{3\pi^2}\Big)^{2/3}
 \frac5{6(\gamma-5)}\Big[\frac{3(\gamma-2)}5+\frac{3\gamma a_1-K_0}T
 \Big]\frac T{\rho_0^{5/3}},\end{equation}
\begin{equation}s_1+2s_2=\Big(\frac2{3\pi^2}\Big)^{2/3}
 \frac3{2(\gamma-5)}\Big[\frac{\gamma-2}3-\frac{\gamma(3J-L)+K_s}T\Big]
 \frac T{\rho_0^{5/3}},\end{equation}
\begin{equation}t_3=\frac{16}{(\gamma-5)(\gamma-3)}\Big[\frac{K_0-15a_1}T
 -\frac95\Big]\frac T{\rho_0^{\gamma/3}},\end{equation}
\begin{equation}x_3=\frac32\frac{T-5(3J-L)-K_s}{K_0-15a_1-\frac95T}
 -\frac12.\end{equation}
Having $s_1$, $s_2$, and $t_3$, $t_0$ can be calculated by 
(\ref{stableSk}). Finally, $x_0$ can be calculated by
\begin{equation}t_0(x_0+\frac12)=\frac2{\gamma-3}\Big[\gamma-2
 -\frac{5\gamma J-(\gamma+2)L+K_s}T\Big]\frac T{\rho_0}.\end{equation}

Myers-Swiatecki interaction parameters can be calculated as:
\begin{equation}\alpha=\frac{K_0-10a_1}T,\end{equation}
\begin{equation}B=\frac5{18}\frac{K_0-6a_1}T,\end{equation}
\begin{equation}\overline\gamma=1-\frac59\frac{K_0-15a_1}T,\end{equation}
\begin{equation}\xi=-\frac{4B(1+\overline\gamma)}{\alpha(4B
 +\overline\gamma)}\Big[1-\frac{5B-\overline\gamma}{B(1
 +\overline\gamma)}\frac JT+\frac{2B-\overline\gamma}{2B(1
 +\overline\gamma)}\frac LT\Big],\end{equation}
\begin{equation}\zeta=\frac13-\frac{2(1+\overline\gamma)}{3(4B
 +\overline\gamma)}\Big[1-\frac3{1+\overline\gamma}\frac{3J-L}T\Big].
\end{equation}

Tondeur interaction parameters are
\begin{equation}a=-\frac35\frac{\gamma-2}{\gamma-3}\Big(1+\frac53
 \frac\gamma{\gamma-2}\frac{a_1}T\Big)\frac T{\rho_0},\end{equation}
\begin{equation}b=\frac3{\gamma-3}\frac1{\rho_0^{\gamma/3}}\Big(a_1
 +\frac T5\Big),\end{equation}
\begin{equation}c=\frac J{\rho_0^{2/3}}.\end{equation}

\begin{table}[h]\caption{The coefficients of volume energy $a_1$, 
symmetry energy$J$, incompressibility $K_0$, density symmetry $L$ and 
symmetry incompressibility $K_s$ calculated from the various Skyrme 
interactions\protect\cite{Brack}(1st to 5th row), Myers-Swiatecki 
interaction\protect\cite{Myers96}(6th row), and Tondeur 
interaction\protect\cite{Tondeur}(7th row), by original parameters, all 
in $MeV$. $\gamma$ is a model parameter in Eqs. (\protect\ref{EoSSk}) 
and (\protect\ref{EoSTo}) for Skyrme and Tondeur interactions 
respectively. $EC$ is the equilibrium criterion calculated from the 
lefthand side of Eq. (\protect\ref{stable0}). As a comparison, the last 
two rows (labeled by CWS) present the result obtained by fitting these
quantities directly to nuclear masses\protect\cite{CWS}. $r_0$ is the 
nuclear radius constant in $fm$.} \vskip 2pt
\begin{tabular}{lcccccccc}%{lcddddddd}
 EOS  &$EC$&$r_0$&$\gamma$&$a_1$&$K_0$& $J$ & $L$ & $K_s$ \\
\hline
 SIII   & 0.00080&1.180& 6  &15.86&355.5&28.16& 9.88&-393.9 \\
 Ska    &-0.00001&1.154& 4  &15.99&263.1&32.91&74.62& -78.45\\
 SkM    & 0.00004&1.142& 7/2&15.77&216.6&30.75&49.34&-148.8 \\
 SkM$^*$& 0.00004&1.142& 7/2&15.77&216.6&30.03&45.78&-155.9 \\
 RATP   & 0.00049&1.143&18/5&16.05&239.6&29.26&32.39&-191.3 \\
 M-S    & 0.00001&1.140&    &16.24&234.4&32.65&49.88&-147.1 \\
 Tondeur& 0.00043&1.145& 4  &15.98&235.8&19.89&39.78& -39.78\\
 CWS    & 0.00000&1.140& 4  &15.98&217.5&28.50&64.32&-101.3 \\
 CWS    & 0.00000&1.140& 5  &16.10&237.9&28.50&63.93&-114.2 \\
\hline\end{tabular}\label{table1}\end{table}

\begin{table}[h]\caption{Input values used to readjust the interaction
parameters, $r_0$ in $fm$, others in $MeV$.} \vskip 2pt
\begin{tabular}{lcccccc}%{lcddddd}
 Force  & $r_0$ & $a_1$ & $K_0$ &  $J$  &  $L$  &  $K_s$  \\
\hline
 Skyrme & 1.140 & 15.97 & 236.07 & 29.25 & 58.50 & -67.92 \\
 M-S    & 1.140 & 15.97 & 236.07 & 29.25 & 58.50 & -67.92 \\
 Tondeur& 1.140 & 15.97 & 236.07 & 29.25 & 58.50 & -58.50 \\
\hline\end{tabular}\label{table2}\end{table}

\begin{table}[h]\caption{Readjusted Skyrme interaction parameters $t_0$,
$t_3$, $x_0$, $s_1$, and $s_2$. Input values are $r_0=1.140fm$,
$a_1=15.97MeV$, $K_0=236.07MeV$, $J=29.25MeV$, $L=58.50MeV$, and
$K_s=-67.92MeV$.} \vskip 2pt
\begin{tabular}{lccccc}%{lddddd}
 Force              & SIII    & Ska     &SkM      & SkM$^*$ & RATP    \\
 \hline
 $\gamma$           & 6       & 4       &7/2      & 7/2     & 18/5    \\
 $t_0(MeVfm^3)$     &-1405.521&-1792.320&-2372.518&-2372.518&-2179.119\\
 $t_3(MeVfm^\gamma)$&-14402.55&12794.56 &12584.33 & 12584.33& 11940.95\\
 $x_0$              & 0.06956 & 0.13735 &0.19759  & 0.19759 & 0.18018 \\
 $x_3$              & 0.38368 & 0.38368 &0.38368  & 0.38368 & 0.38368 \\
 $s_1(MeVfm^5)$     & 642.825 & -42.389 &71.813   &  71.813 & 55.499  \\
 $s_2(MeVfm^5)$     &-473.186 &  84.802  & -8.196   & -8.196 & 5.090  \\
 \hline\end{tabular}\label{table3}\end{table}

\begin{table}[h]\caption{Critical point $(\rho_c,\delta_c)$ predicted by
Skyrme, Myers-Swiatecki, and Tondeur interactions respectively. $r_0$ in 
$fm$, $\rho_c$ in $fm^{-3}$, and $e_c$ in $MeV$. For each item, the 
first line is given by original interaction parameters, the second line 
by readjusted parameters shown in Table 3 for Skyrme interactions while 
by Eqs. (\protect\ref{paraMS}) and (\protect\ref{paraTo}) for
Myers-Swiatecki and Tondeur interactions respectively.}\vskip 2pt
\begin{tabular}{lccccccc}%{lcdddddd}
           & SIII & Ska  & SkM &SkM$^*$& RATP & M-S  &Tondeur\\
\hline
 $r_0$     & 1.180& 1.154& 1.142& 1.142& 1.143& 1.140& 1.145 \\
           & 1.140& 1.140& 1.140& 1.140& 1.140& 1.140& 1.140 \\
 $\delta_c$& 0.8385& 0.8647& 0.8390& 0.8421& 0.8303& 0.8213& 0.8732 \\
           & 0.8772& 0.8980& 0.8908& 0.8908& 0.8920& 0.8988& 0.7697 \\
 $\rho_c$  &0.07173&0.02416&0.02345&0.02420&0.03892&0.03039&0.03081 \\
           &0.02732&0.02969&0.02825&0.02825&0.02851&0.02643&0.03131 \\
 $e_c$     & 3.9019& 1.5852& 1.2572& 1.2814& 1.9898& 1.1031& 2.6142 \\
           & 1.8894& 1.8505& 1.7025& 1.7025& 1.7311& 1.1280& 2.6304 \\
\hline\end{tabular}\label{table4}\end{table}

\begin{figure*}\caption{(a) Normalized energy per nucleon $e/a_1$ versus 
relative nucleon density $\rho/\rho_0$ calculated by original interaction 
parameters at $\delta=0$, (b) $e$ versus $\rho/\rho_0$ calculated by 
readjusted interaction parameters at $\delta=0$ of the various Skyrme 
interactions (solid lines), Myers-Swiatecki interaction (dot-dashed line), 
and Tondeur interaction (dashed line). The solid lines in the righthand 
side of the plot in (a) from top to bottom correspond to SIII, Ska, RATP, 
SkM, and SkM* interactions respectively. The difference between SkM and 
SkM* is negligible and Myers-Swiatecki is almost coincident with RATP.}
\label{Figure1}\end{figure*}

\begin{figure*}\caption{(a) Normalized nuclear incompressibility $K/K_0$ 
versus relative nucleon density $\rho/\rho_0$ calculated by original 
interaction parameters at $\delta=0$, (b) $K$ versus $\rho/\rho_0$ 
calculated by readjusted interaction parameters at $\delta=0$ of the 
various Skyrme interactions (solid lines), Myers-Swiatecki interaction 
(dot-dashed line), and Tondeur interaction (dashed line). On the 
righthand side of the plot in (a) the solid lines from top to bottom 
correspond to SIII, Ska, RATP, SkM, and SkM* interactions respectively. 
The difference between SkM and SkM* is negligible. In (b), the solid lines 
from top to bottom are due to SkM, SkM*, RATP, Ska, and SIII, where SkM* 
is identical to SkM. Tondeur is coincident with Ska; SkM and RATP almost 
overlap.}\label{Figure2}\end{figure*}

\begin{figure*}\caption{(a) Normalized symmetry energy $J(\rho)/J$ versus 
the relative nucleon density $\rho/\rho_0$ calculated by original 
interaction parameters, (b) $J(\rho)$ versus $\rho/\rho_0$ calculated by 
readjusted interaction parameters of the various Skyrme interactions 
(solid lines), Myers-Swiatecki interaction (dot-dashed line), and Tondeur 
interaction (dashed line). On the righthand side of the plot, in (a), the 
solid lines, from top to bottom, correspond to Ska, SkM, SkM*, RATP, and 
SIII interactions respectively. The difference between SkM and SkM* for 
$\rho/\rho_0<1.3$ is negligible. In addition, the Myers-swiatecki's is 
almost coincident with that of SkM's. In (b), the difference among these 
curves is negligible.}\label{Figure3}\end{figure*}

\begin{figure*}\caption{Equilibrium density $\rho_m$ as function of the
relative neutron excess $\delta$, calculated by (a) original parameters 
and (b) readjusted parameters of the various Skyrme interactions (solid 
lines), Myers-Swiatecki interaction (dot-dashed line), and Tondeur 
interaction (dashed line). The solid lines, in the middle range of $\delta$ 
from top to bottom in (a) correspond to SIII, RATP, SkM, SkM*, and Ska 
interactions respectively, the difference between SkM and SkM* is very 
small. In (b) the solid lines from top to bottom correspond to Ska, RATP, 
SkM, SkM*, and SIII interactions respectively. SkM and SkM* are the same 
whereas RATP, SkM and Myers-Swiatecki almost overlap.}
\label{Figure4}\end{figure*}

\begin{figure*}\caption{Equilibrium energy per nucleon $e_m$ as function 
of the relative neutron excess $\delta$, calculated by (a) original 
parameters and (b) readjusted parameters of the various Skyrme interactions 
(solid lines), Myers-Swiatecki interaction (dot-dashed line), and Tondeur 
interaction (dashed line). The solid lines, in the righthand side, in (a) 
correspond to RATP, SkM, SkM*, SIII, and Ska interactions from top to
bottom respectively. SIII, Ska and Tondeur almost overlap, whereas the 
difference between SkM and SkM* is negligible. In (b) the solid lines from 
top to bottom correspond to SIII, SkM, SkM*, RATP, and Ska interactions, 
respectively, where SkM and SkM* are the same. SkM, RATP and Ska are almost 
coincident.}\label{Figure5}\end{figure*}

\begin{figure*}\caption{Equilibrium incompressibility $K_m$ as function 
of the relative neutron excess $\delta$, calculated by (a) original 
parameters and (b) readjusted parameters of the various Skyrme interactions 
(solid lines), Myers-Swiatecki interaction (dot-dashed line), and Tondeur 
interaction (dashed line). The solid lines in (a) correspond to SIII, RATP, 
Ska, SkM*, and SkM interactions from top to bottom in the middle range of
$\delta$, respectively. Ska, SkM* and SkM almost overlap. In (b) the solid 
lines from top to bottom correspond to Ska, RATP, SkM, SkM*, and SIII 
interactions respectively; SkM and SkM* are the same.}
\label{Figure6}\end{figure*}

\begin{figure*}\caption{Neutron matter EOS $e(\rho,1)$ versus nucleon 
density $\rho$, calculated by (a) original parameters and (b) readjusted 
parameters of the various Skyrme interactions (solid lines), Myers-Swiatecki 
interaction (dot-dashed line), and Tondeur interaction (dashed line). Black 
diamonds denote the data taken from Ref. \protect\cite{Friedman}. The 
solid lines, in the right hand side of (a), correspond to Ska, SkM, SkM*, 
RATP, and SIII from top to bottom. In (b), the solid lines, from top to 
bottom, correspond to SIII, SkM, SkM*, RATP, and Ska, where SkM and SkM* 
are identical.}\label{Figure7}\end{figure*}

\begin{figure*}\caption{Symmetric nuclear matter EOS $e(\rho,0)$, calculated 
by (a) original parameters and (b) readjusted parameters of the various 
Skyrme interactions (solid lines), Myers-Swiatecki interaction (dot-dashed 
line), and Tondeur interaction (dashed line). The full dots stand for the 
results taken from Ref. \protect\cite{Wiringa}, the crosses denote the 
results taken from Ref. \protect\cite{Akmal}. The solid lines, in the right 
hand side of (a), correspond to SIII, Ska, RATP, SkM, and SkM*, from top to 
bottom, where SkM and SkM* are identical. In (b), the solid lines, from top 
to bottom, correspond to SkM, SkM*, RATP, Ska, and SIII, where SkM and SkM* 
are identical. The Tondeur's is very close to the Ska's.}
\label{Figure8}\end{figure*}

\begin{figure*}\caption{Neutron matter EOS $e(\rho,1)$, calculated by (a) 
original parameters and (b) readjusted parameters of the various Skyrme 
interactions (solid lines), Myers-Swiatecki interaction (dot-dashed line), 
and Tondeur interaction (dashed line). The full dots stand for the results 
taken from Ref. \protect\cite{Wiringa}, the crosses denote the results 
taken from Ref. \protect\cite{Akmal}. The solid lines, in the right hand 
side of (a), correspond to Ska, RATP, SkM, SkM*, and SIII, from top to 
bottom. In (b), the solid lines, from top to bottom, correspond to SIII, 
SkM, SkM*, RATP, and Ska, where SkM and SkM* are identical.}
\label{Figure9}\end{figure*}

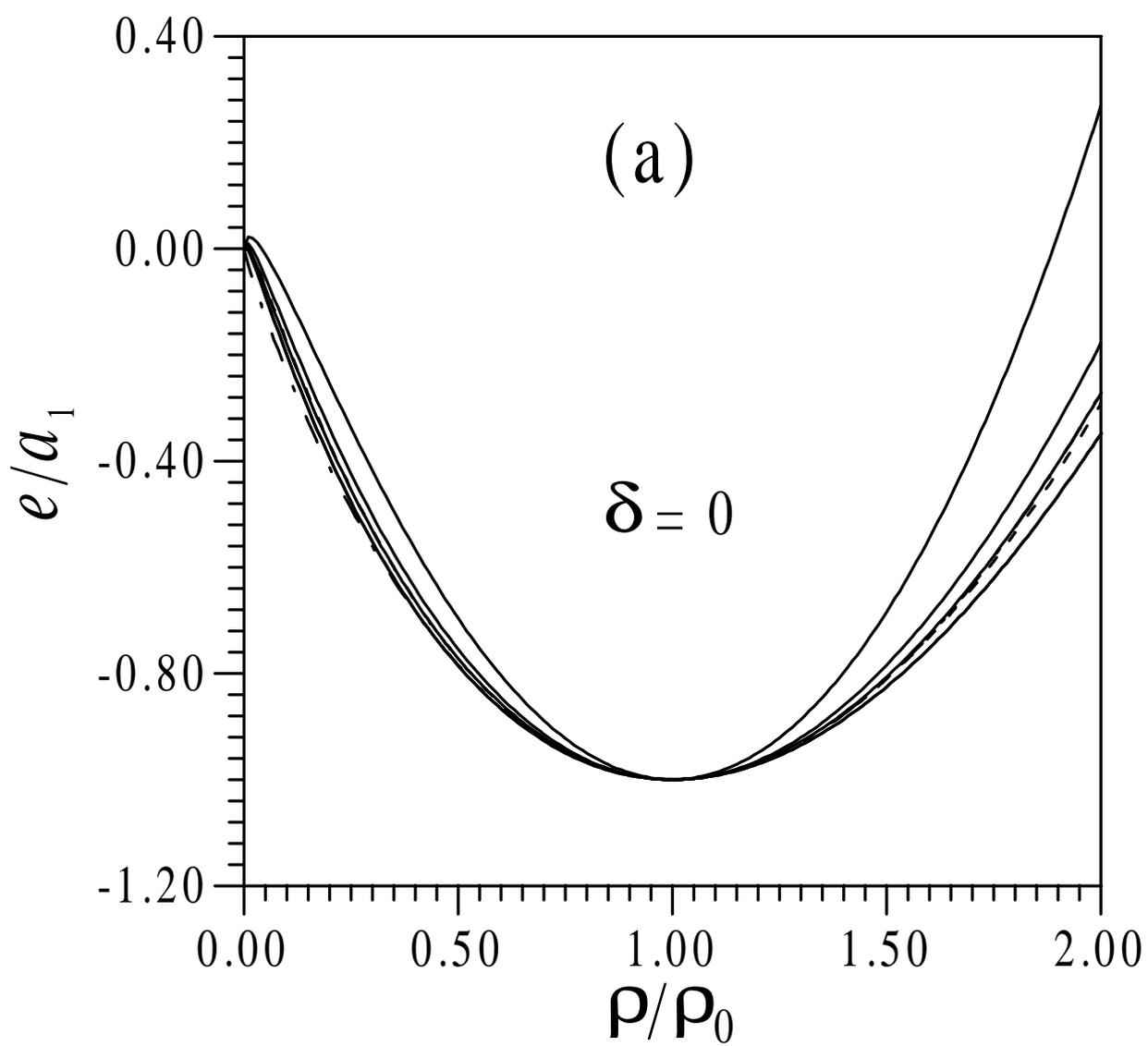
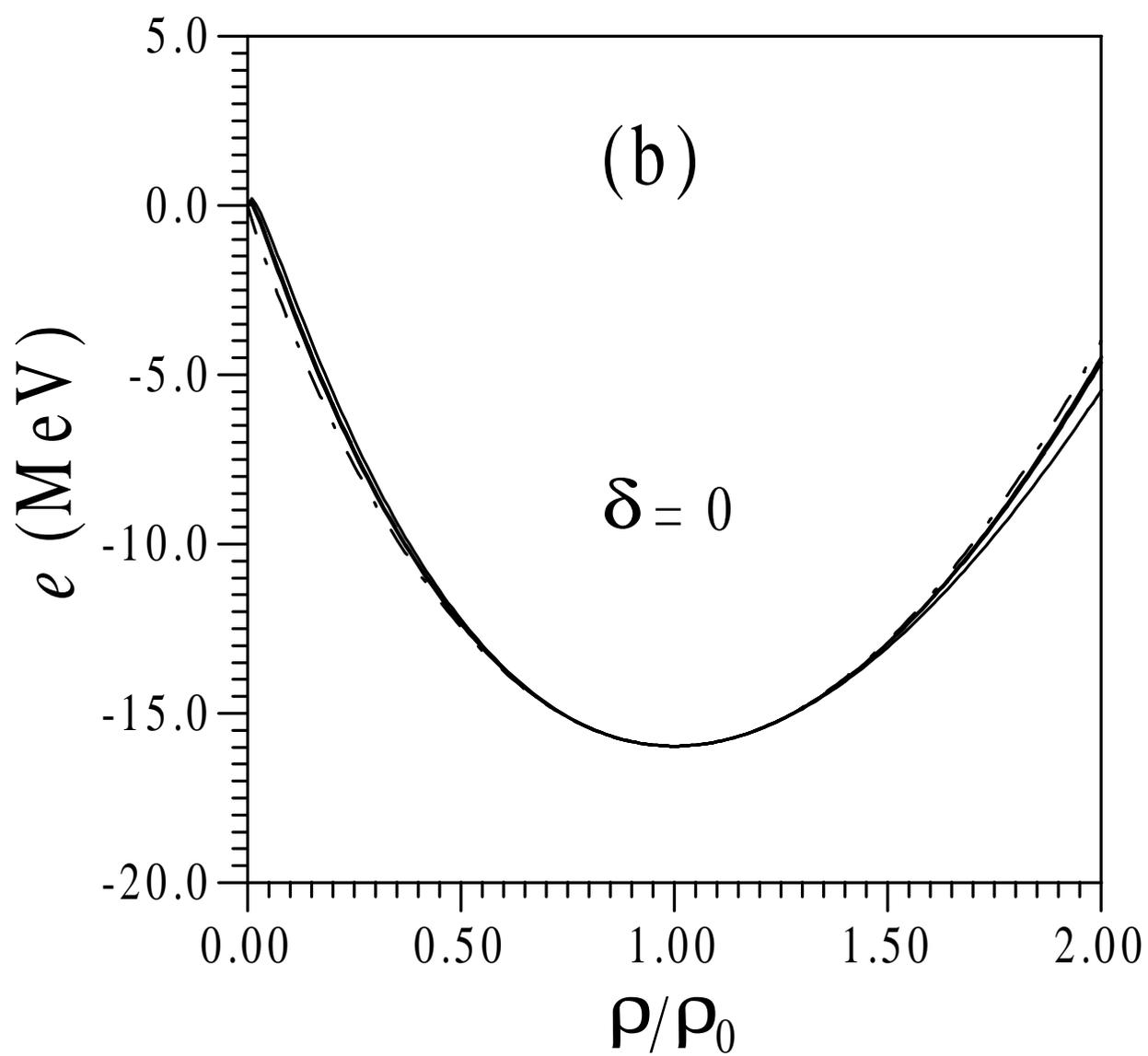
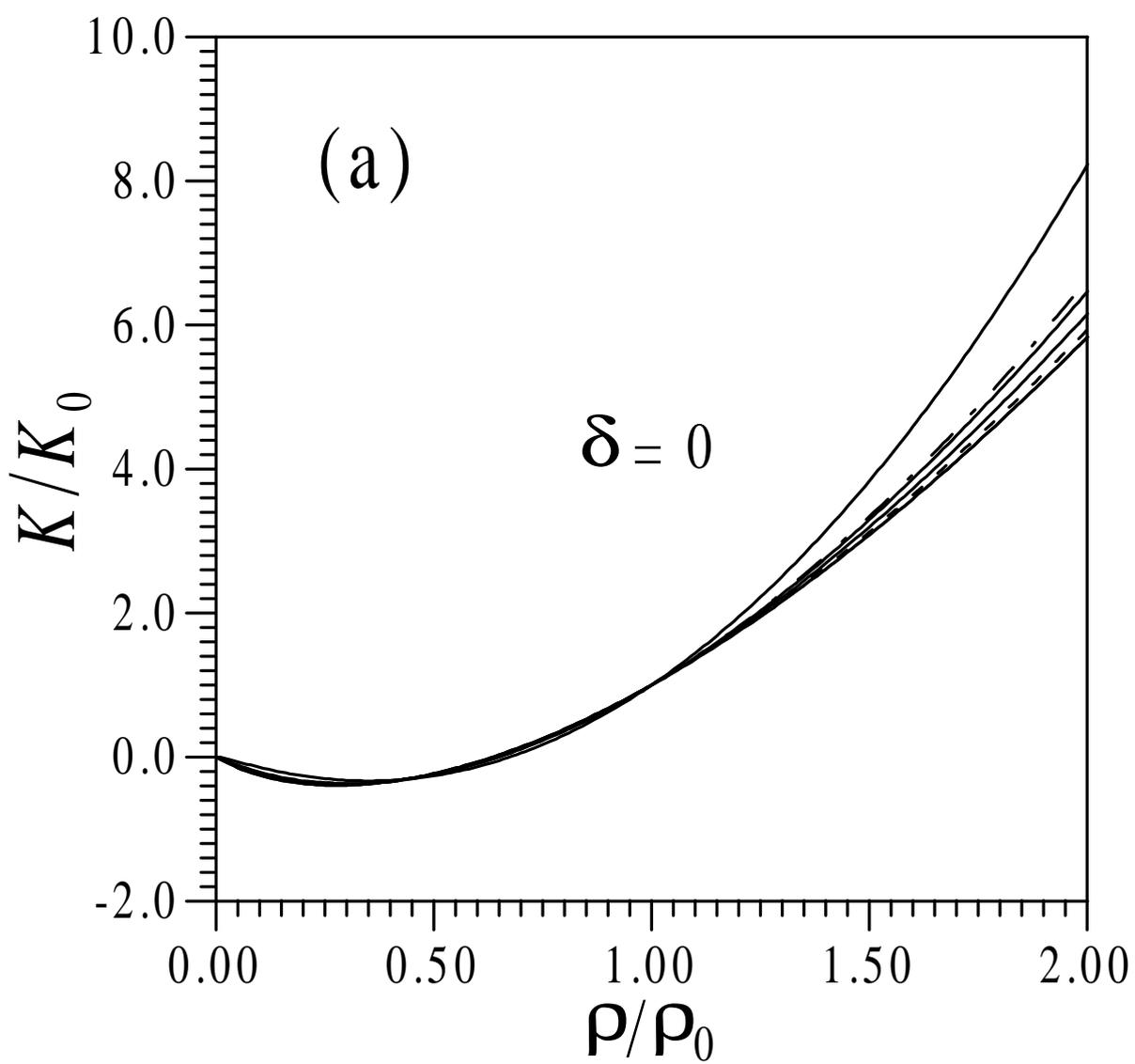
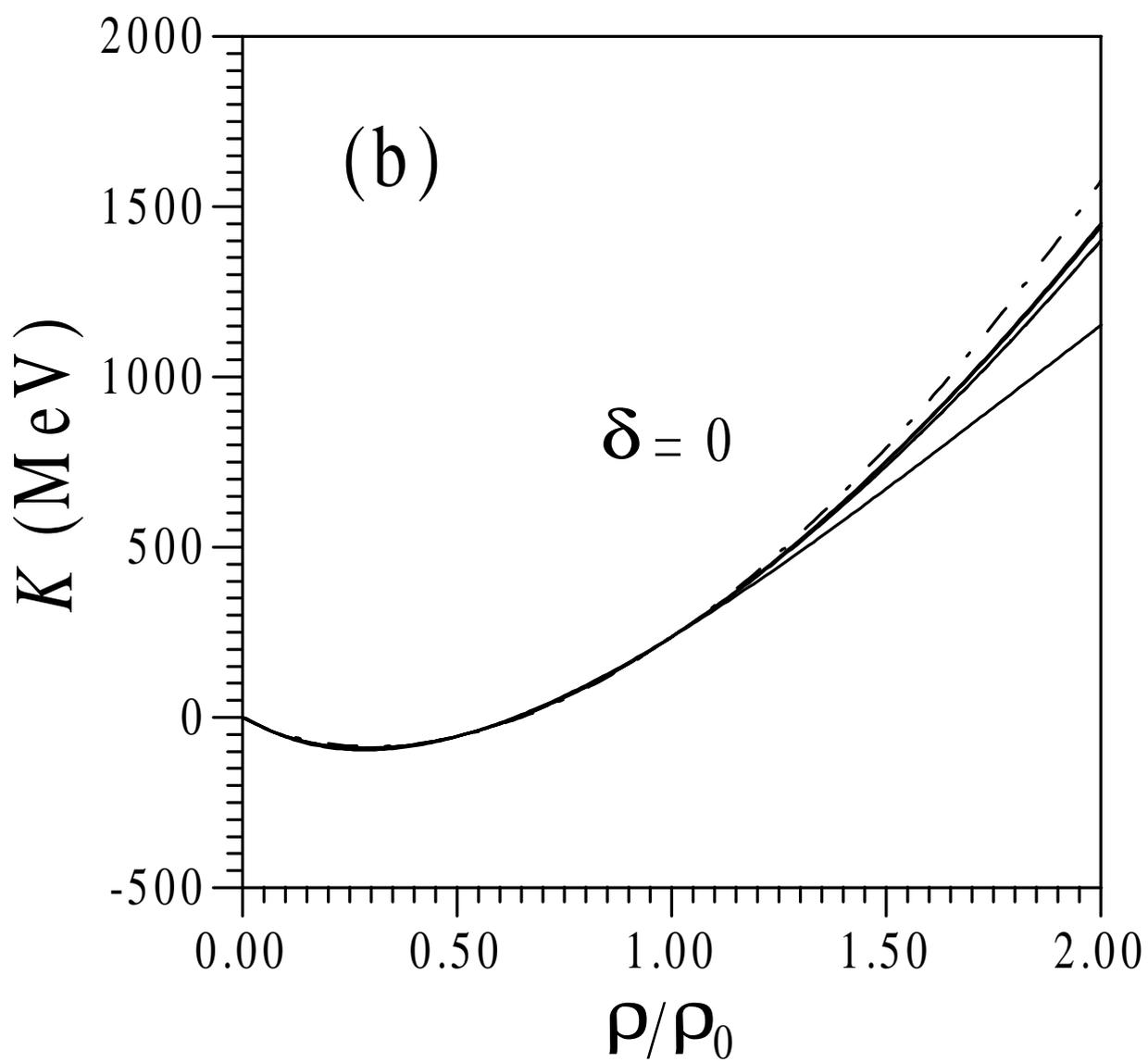
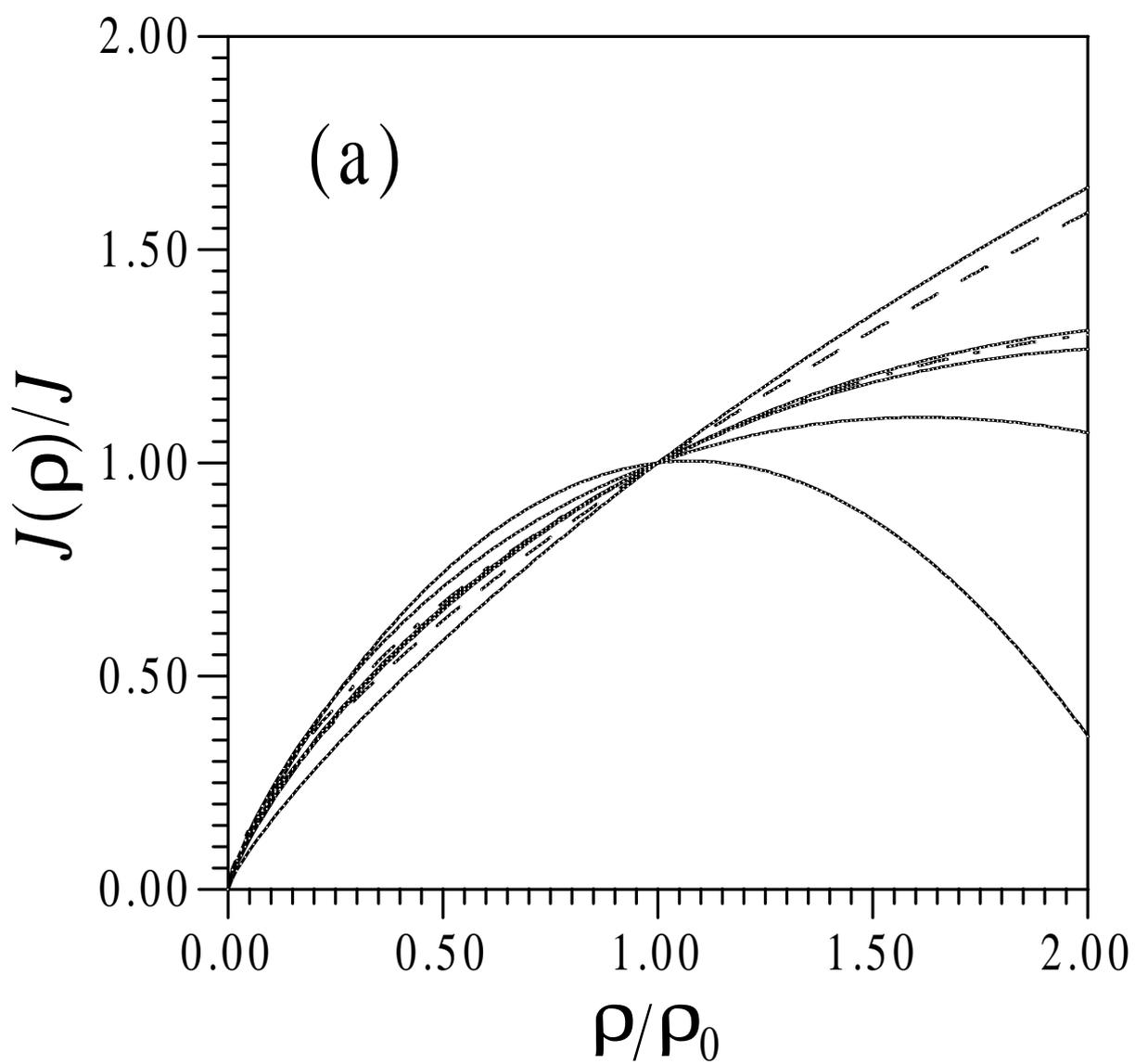
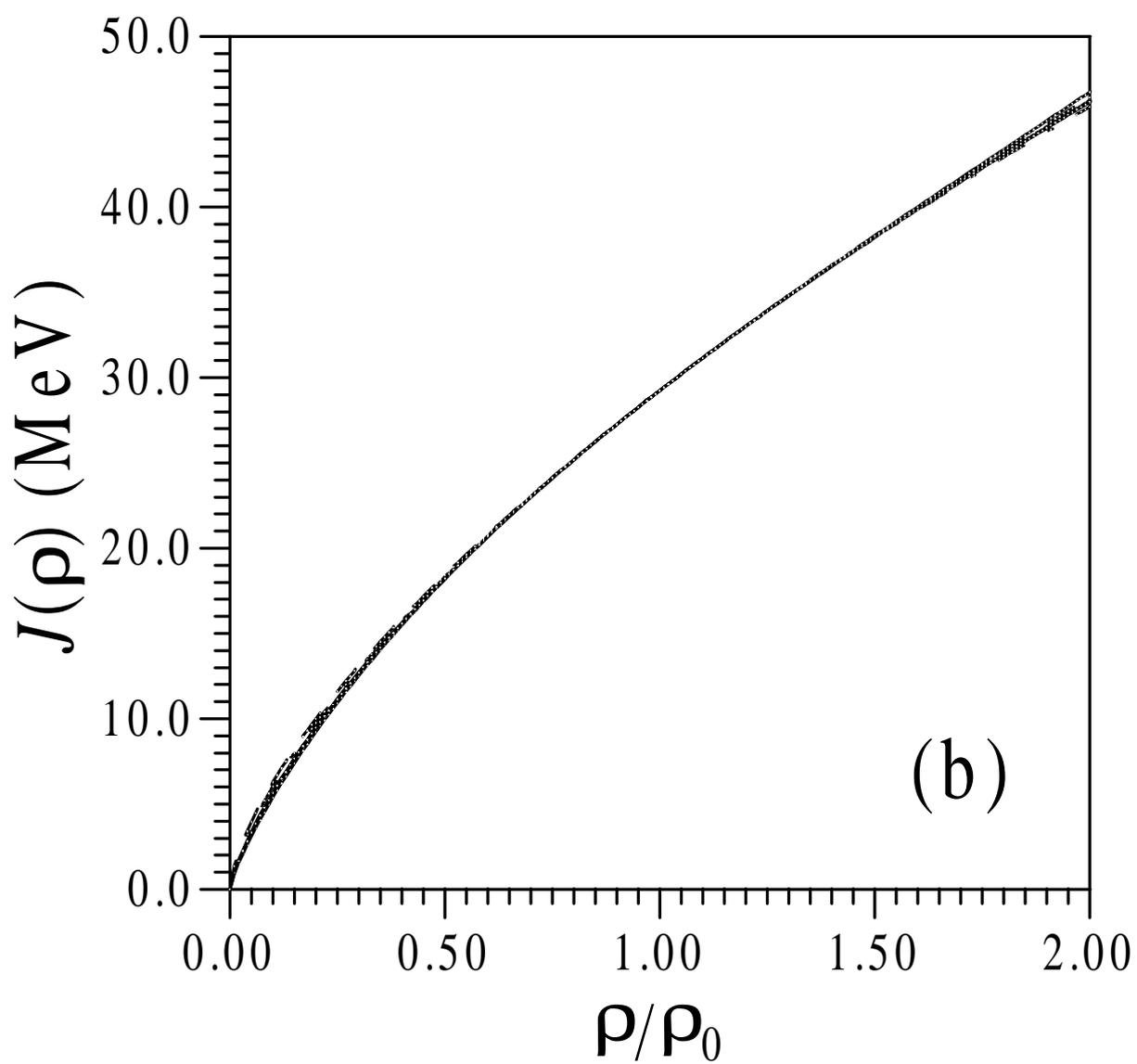
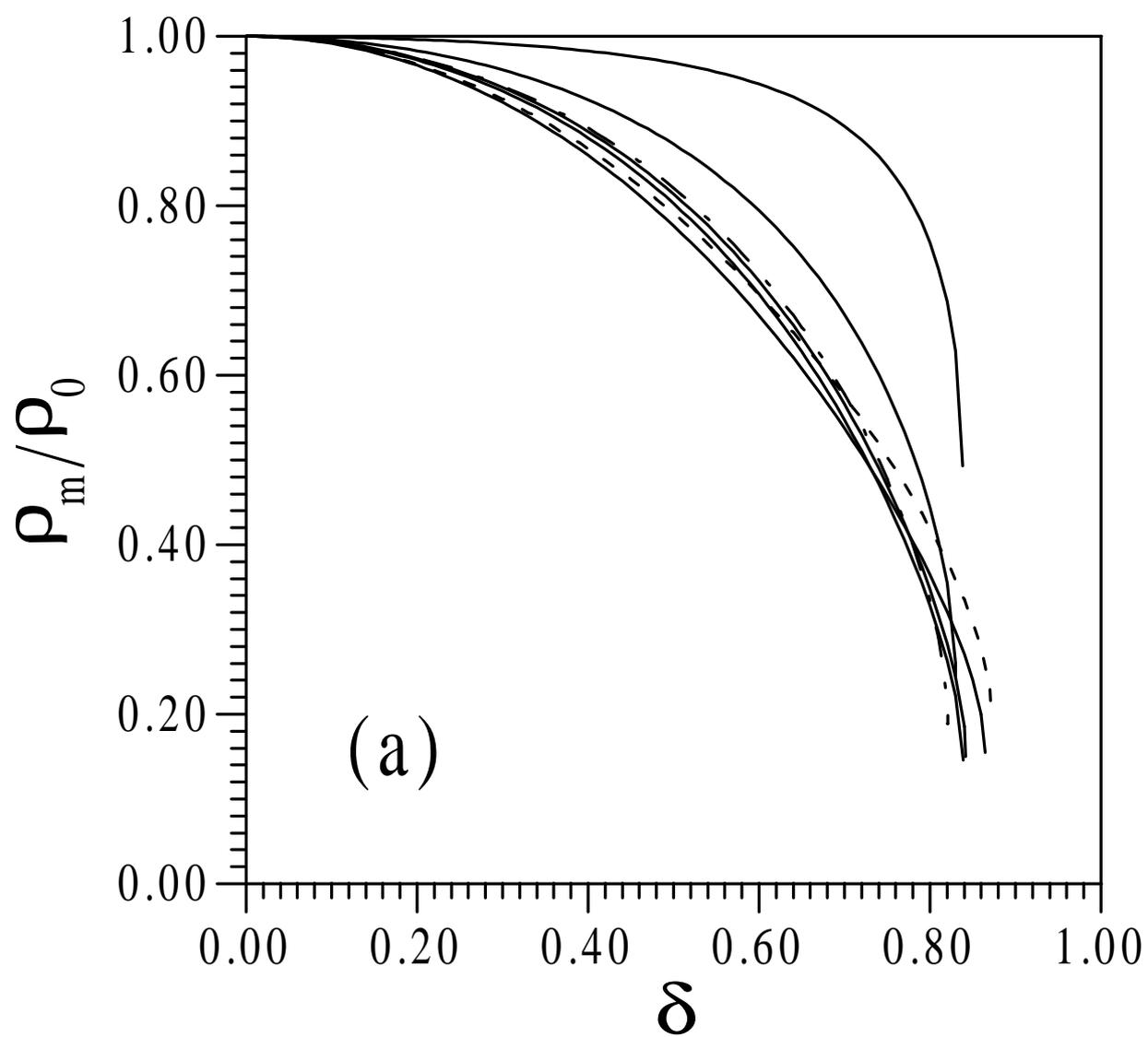
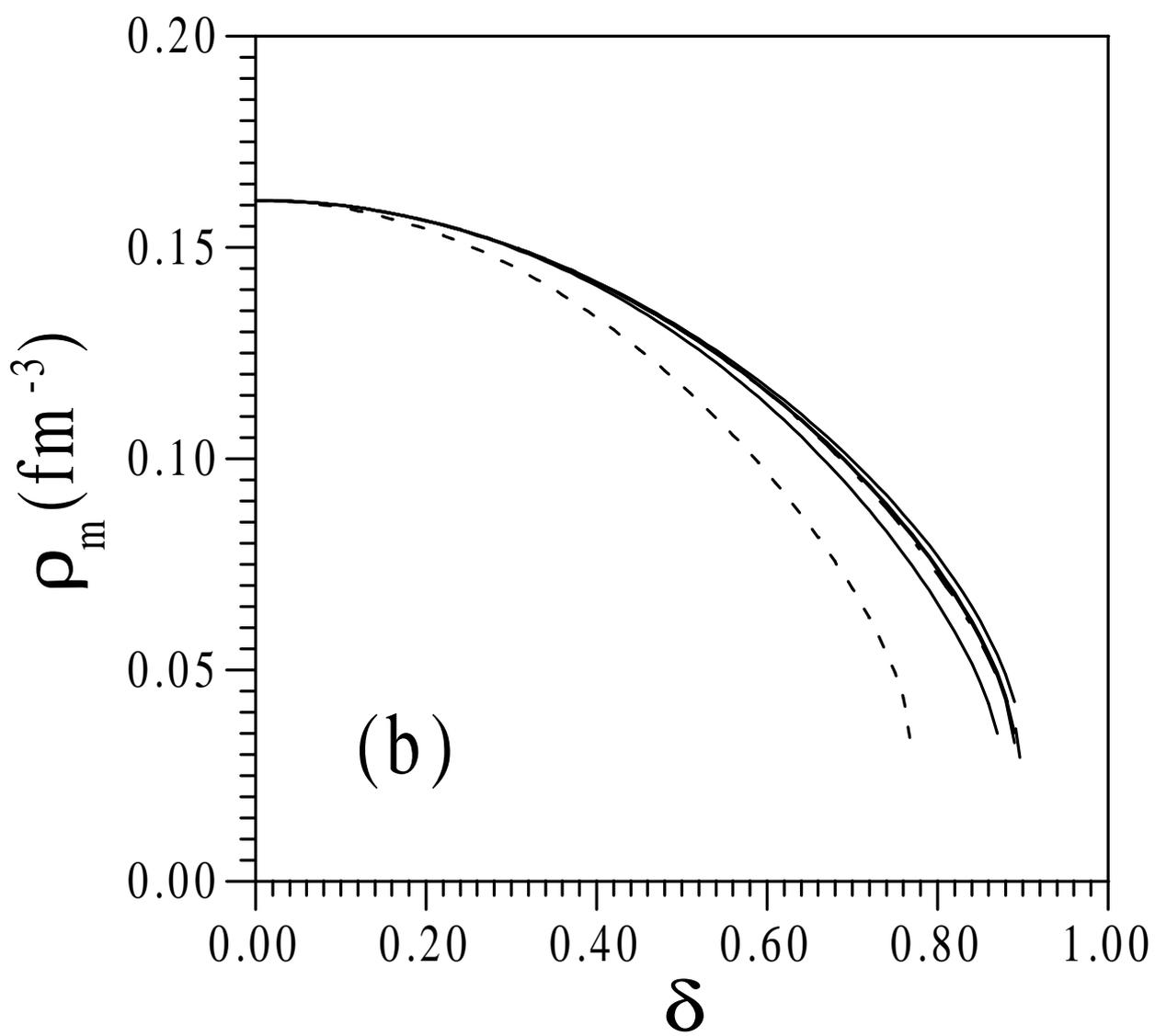
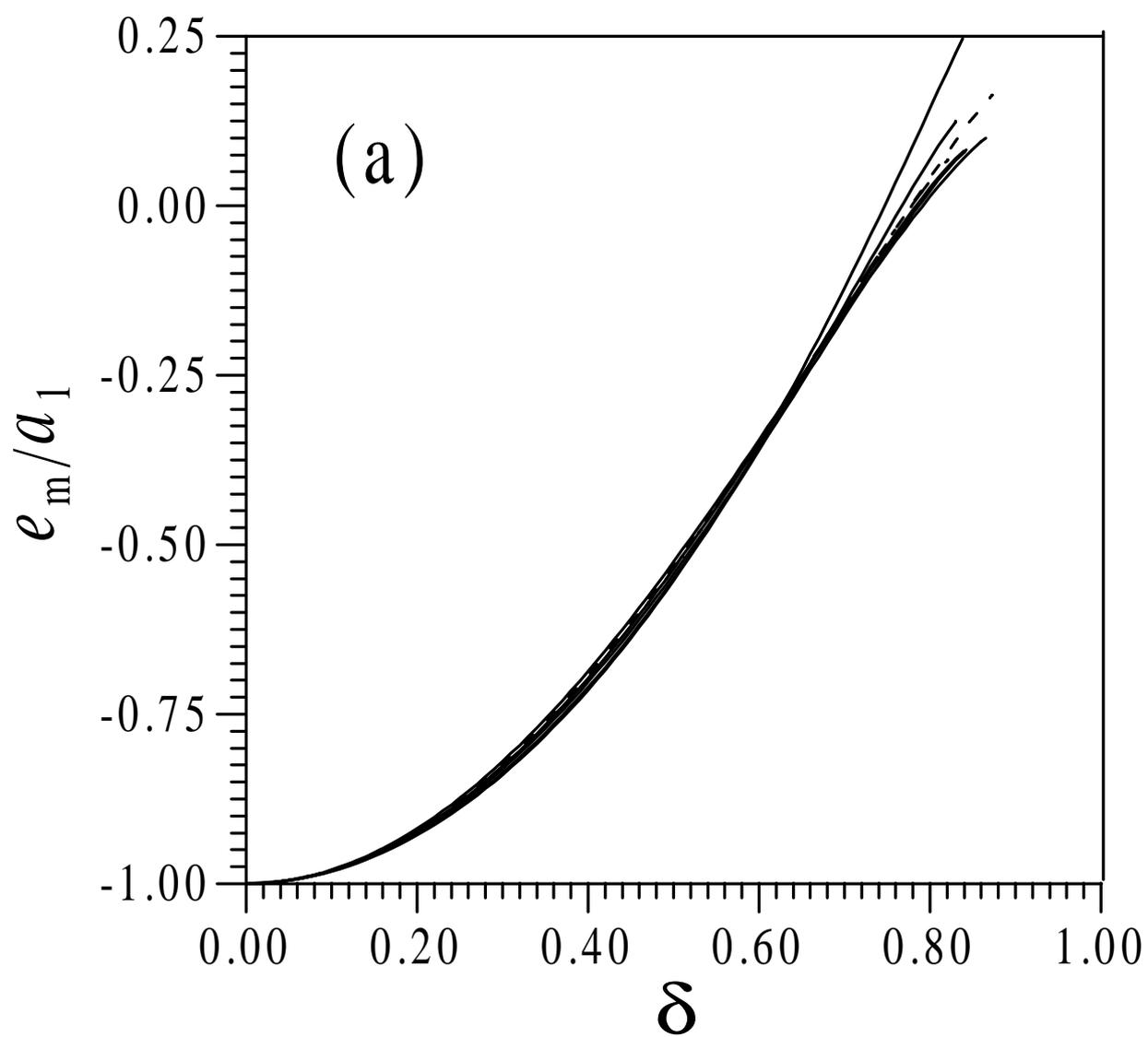
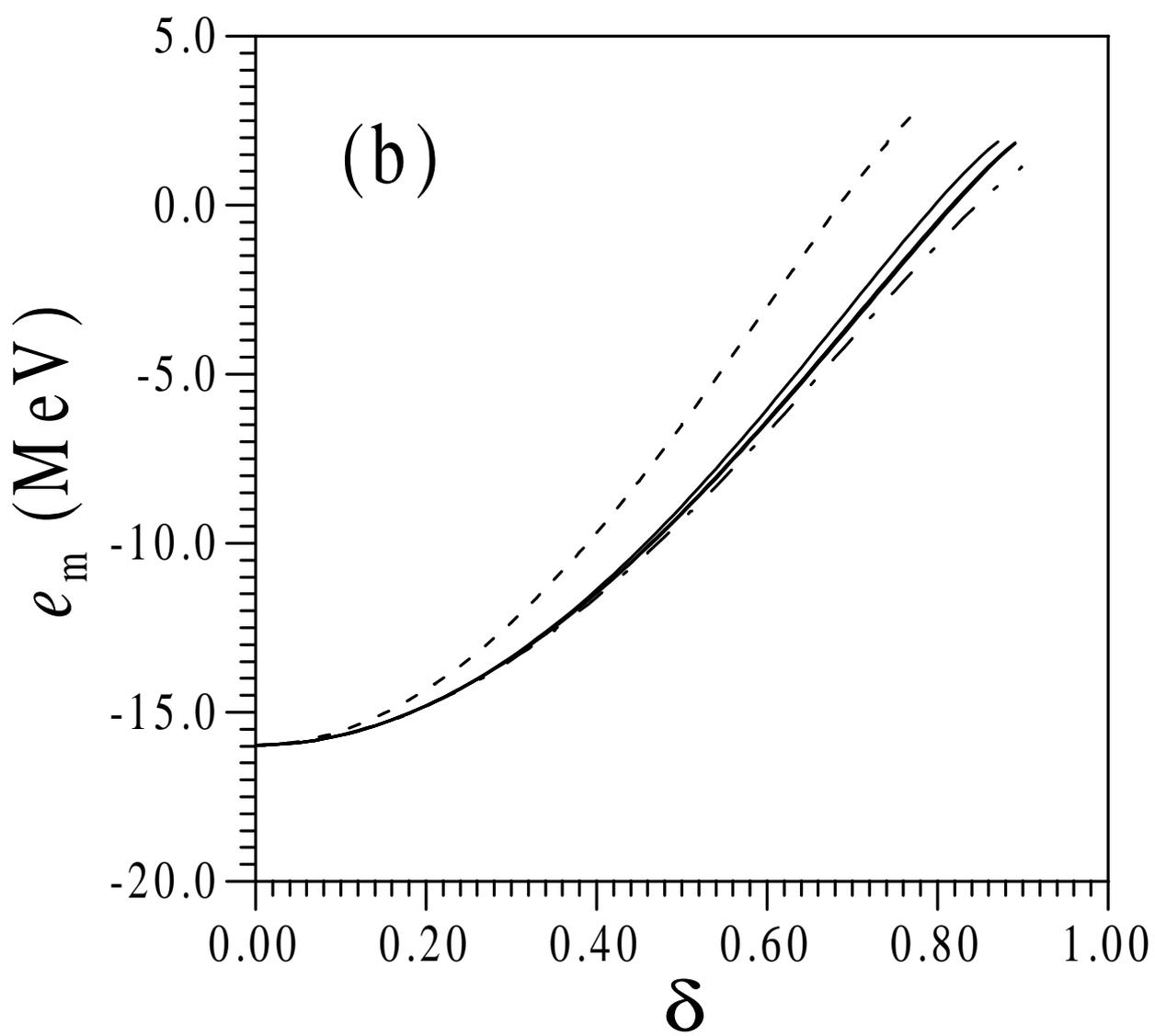
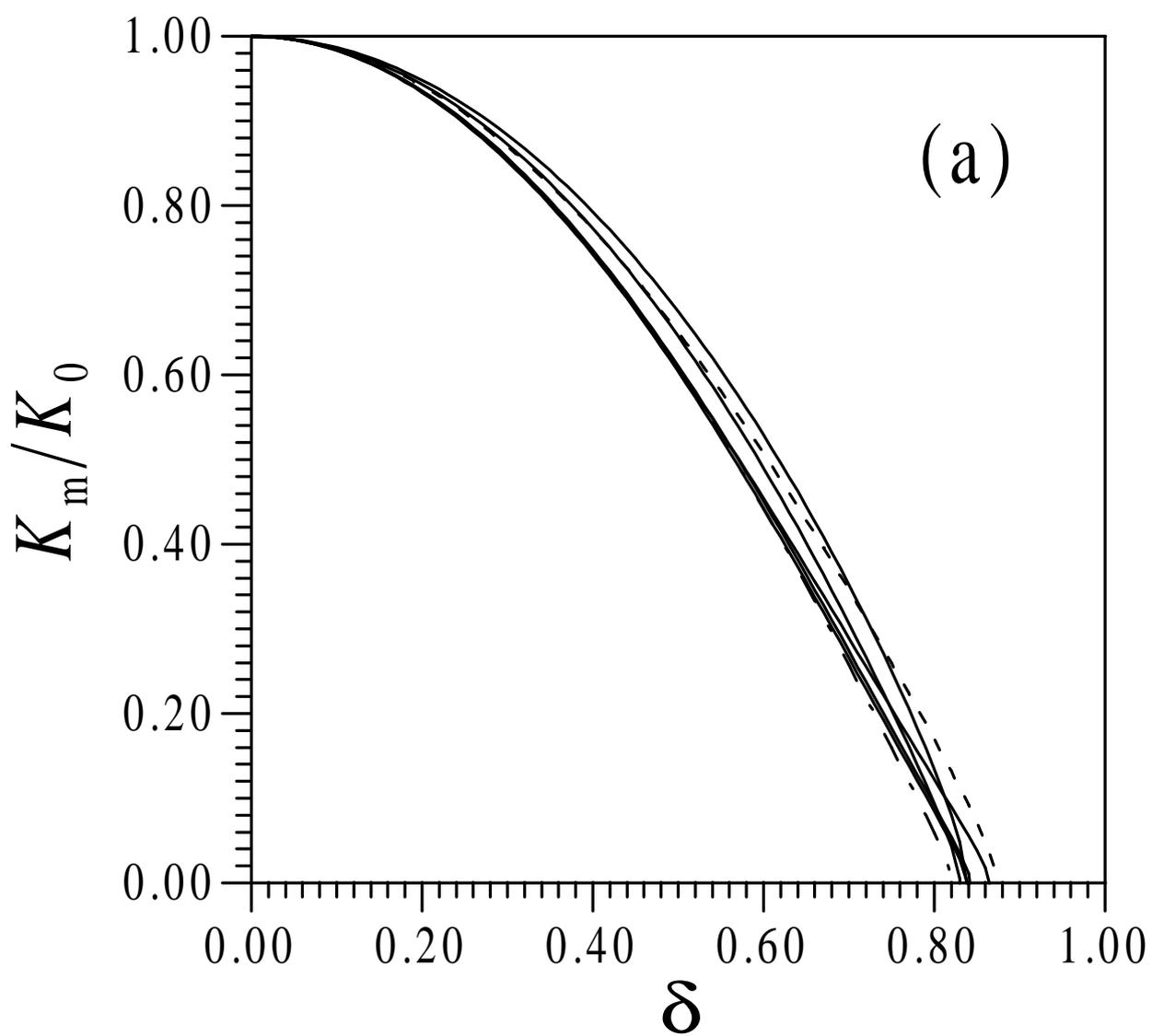
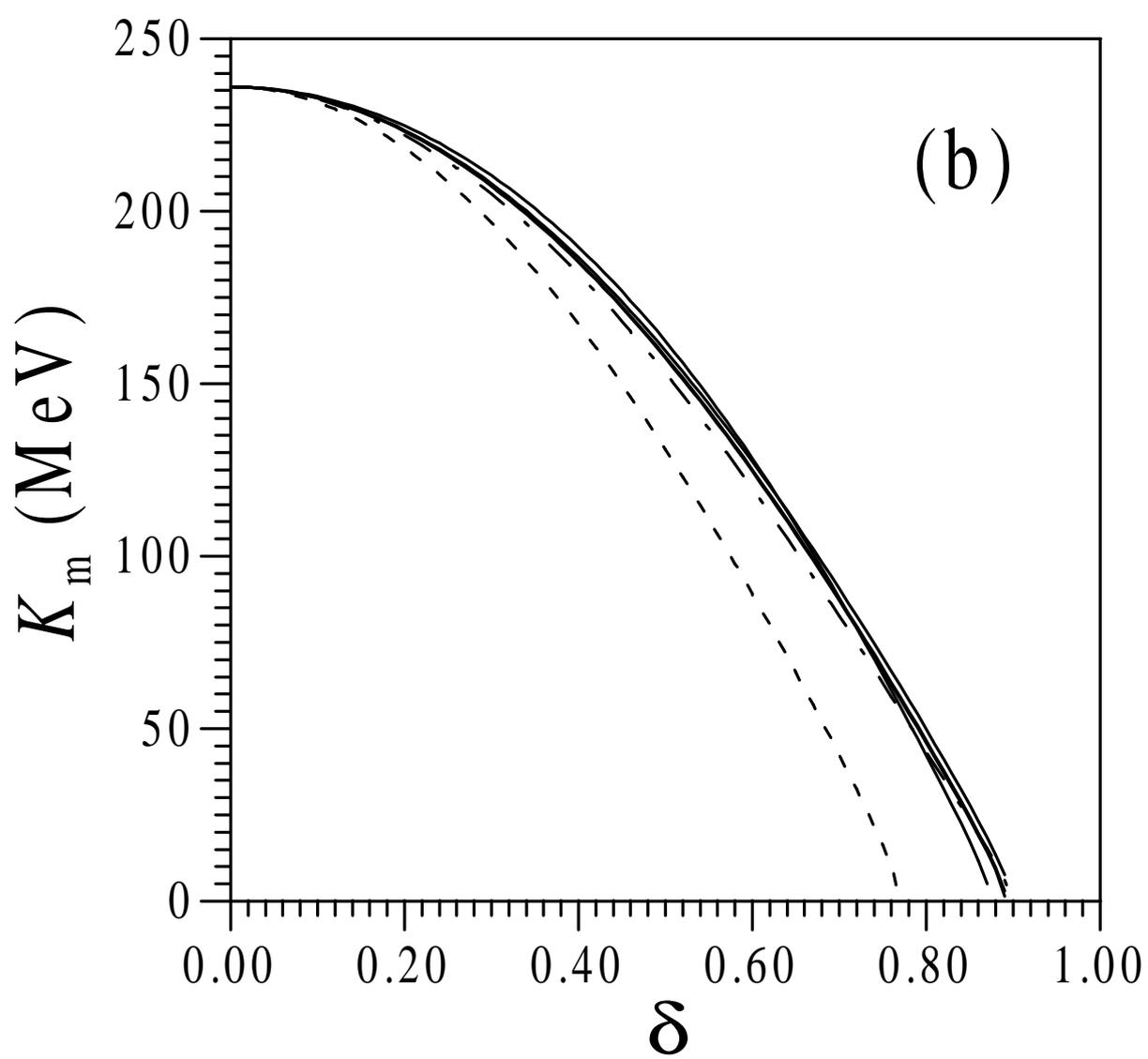
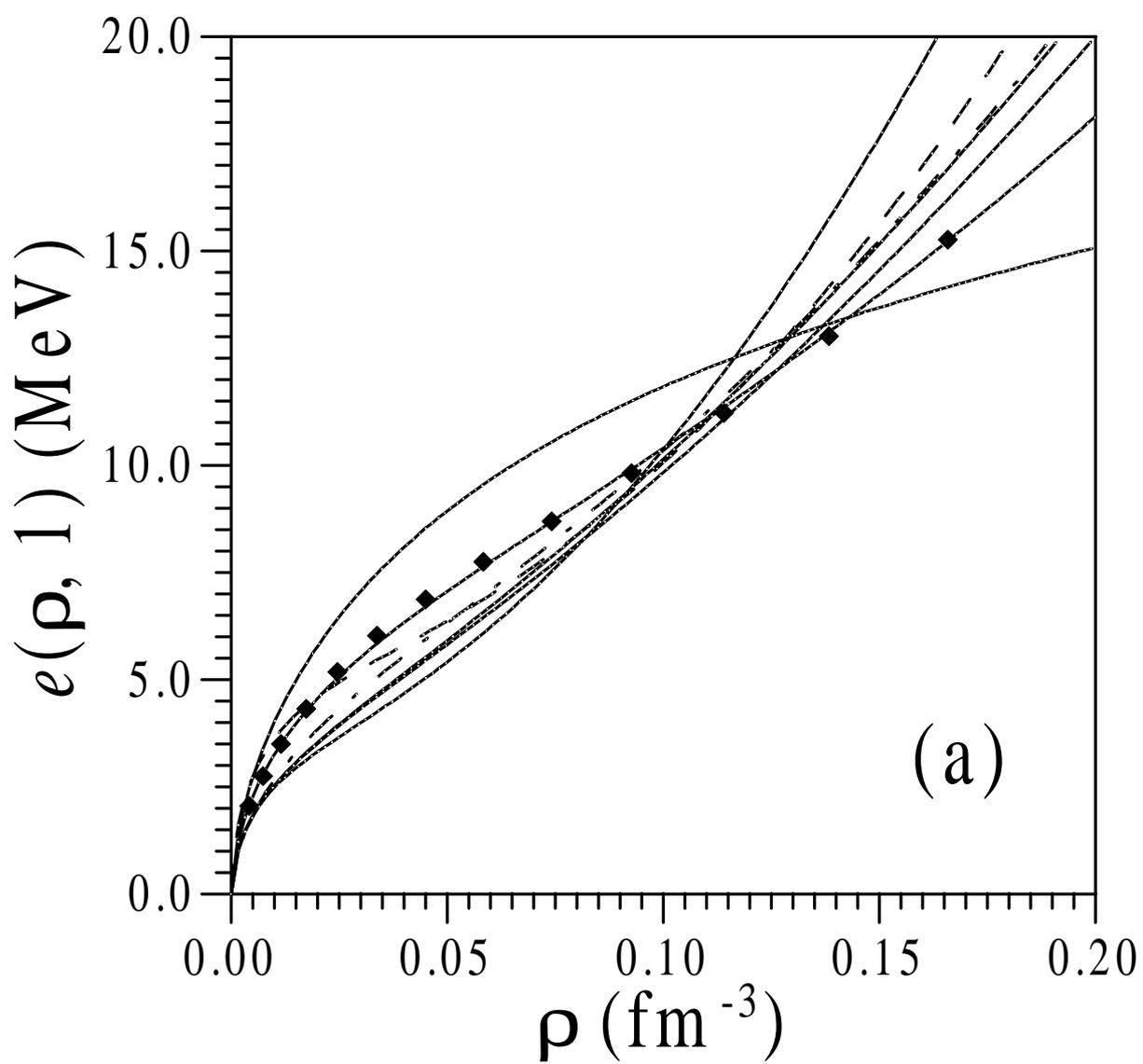
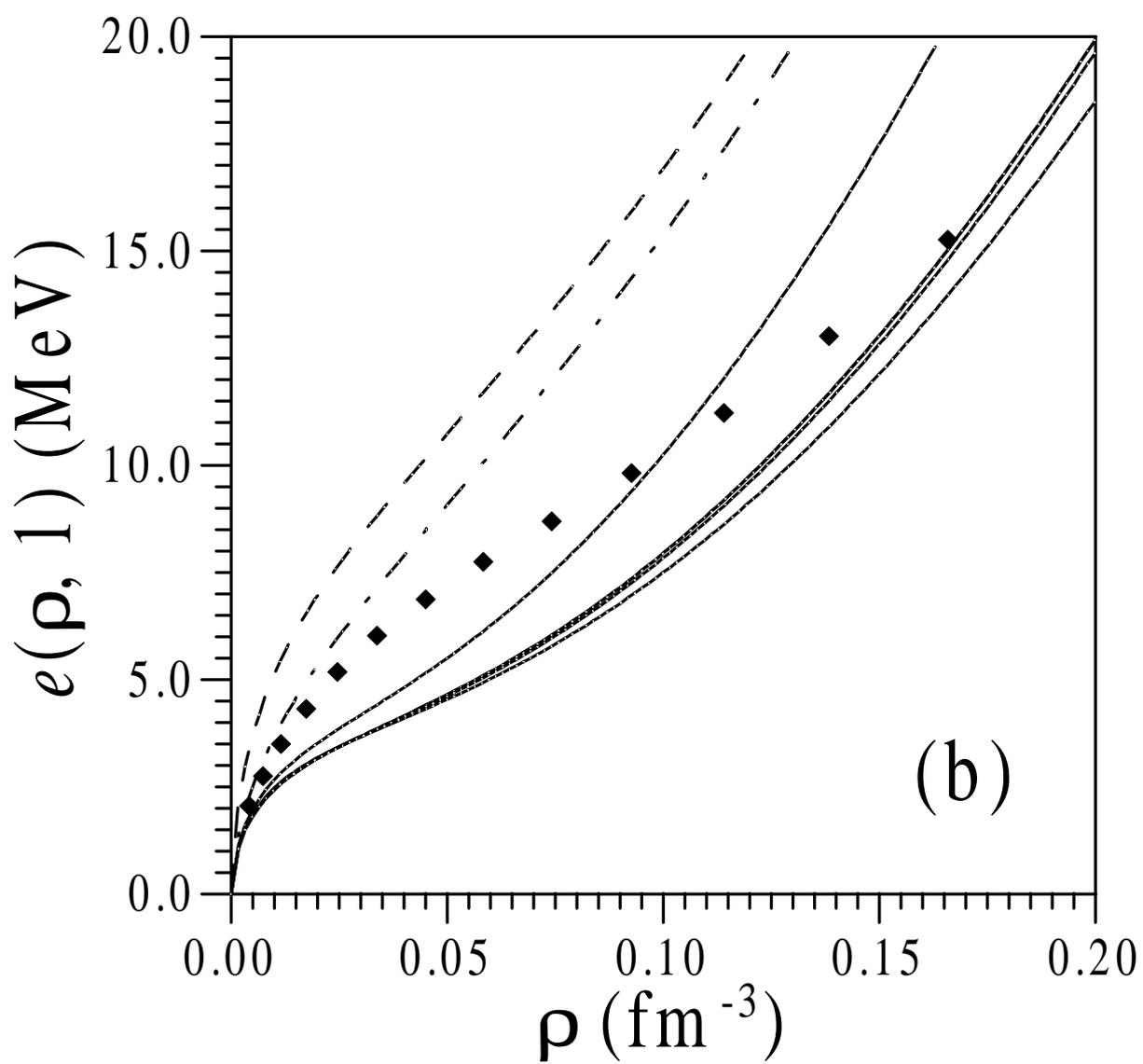
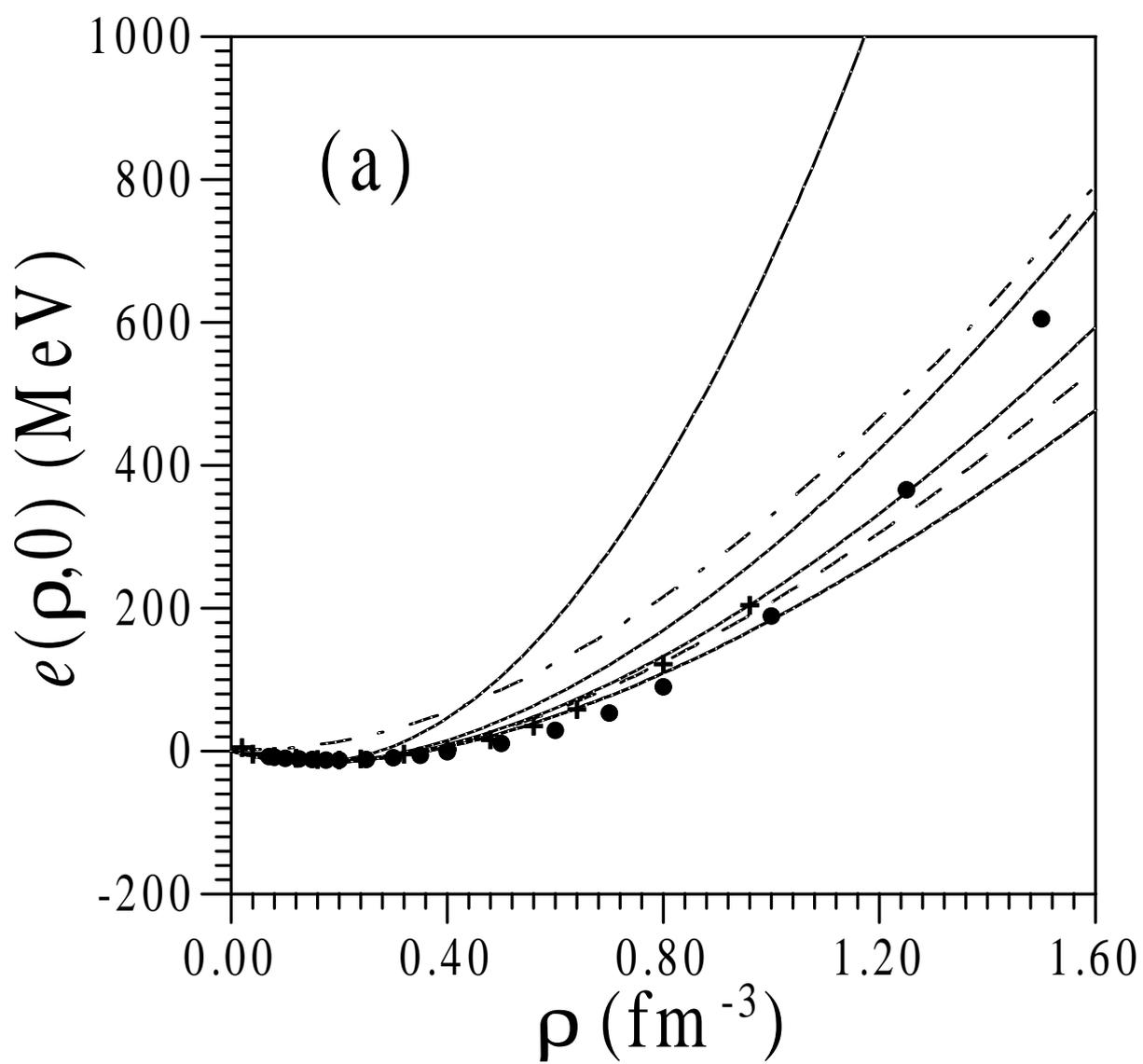
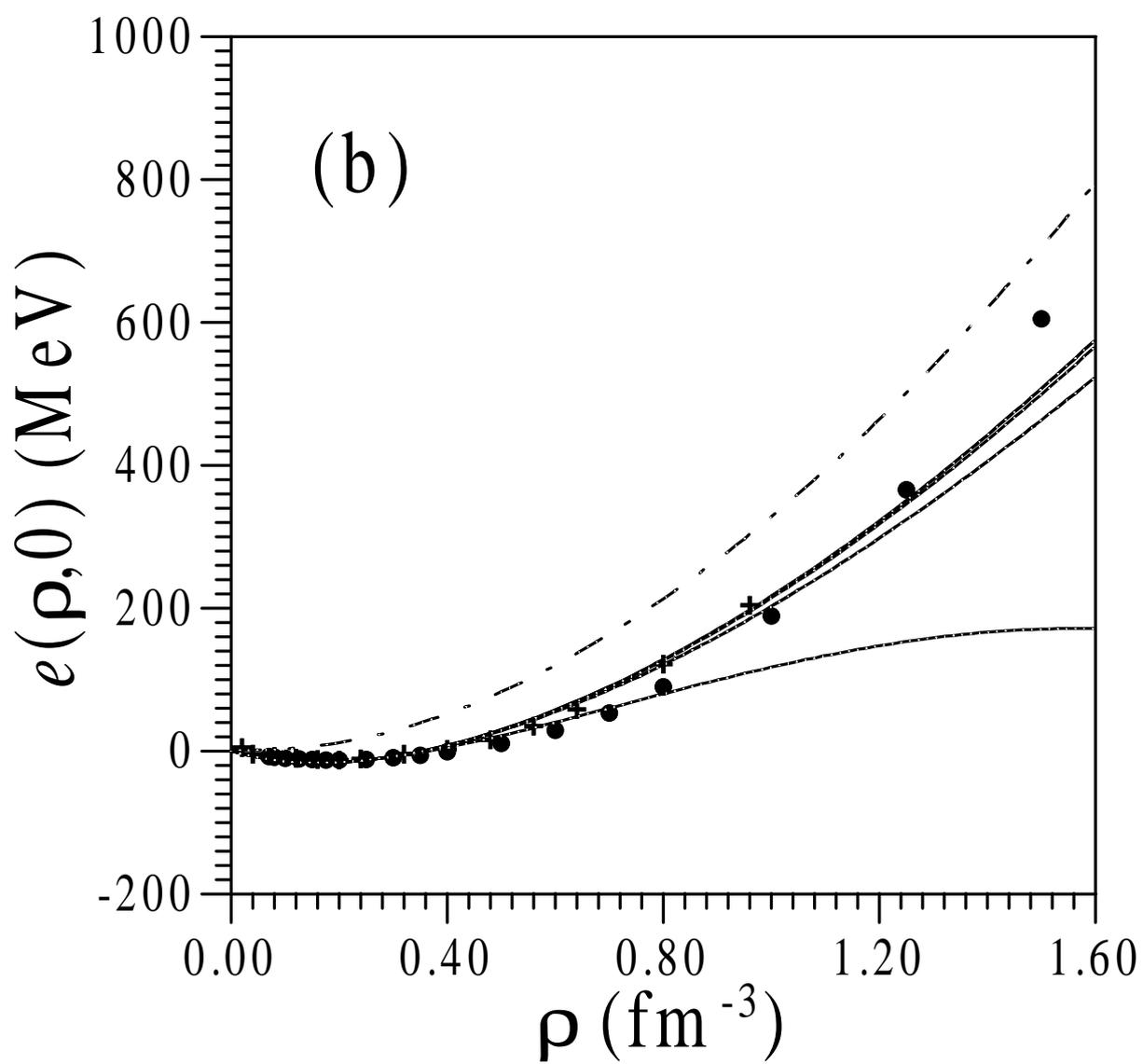
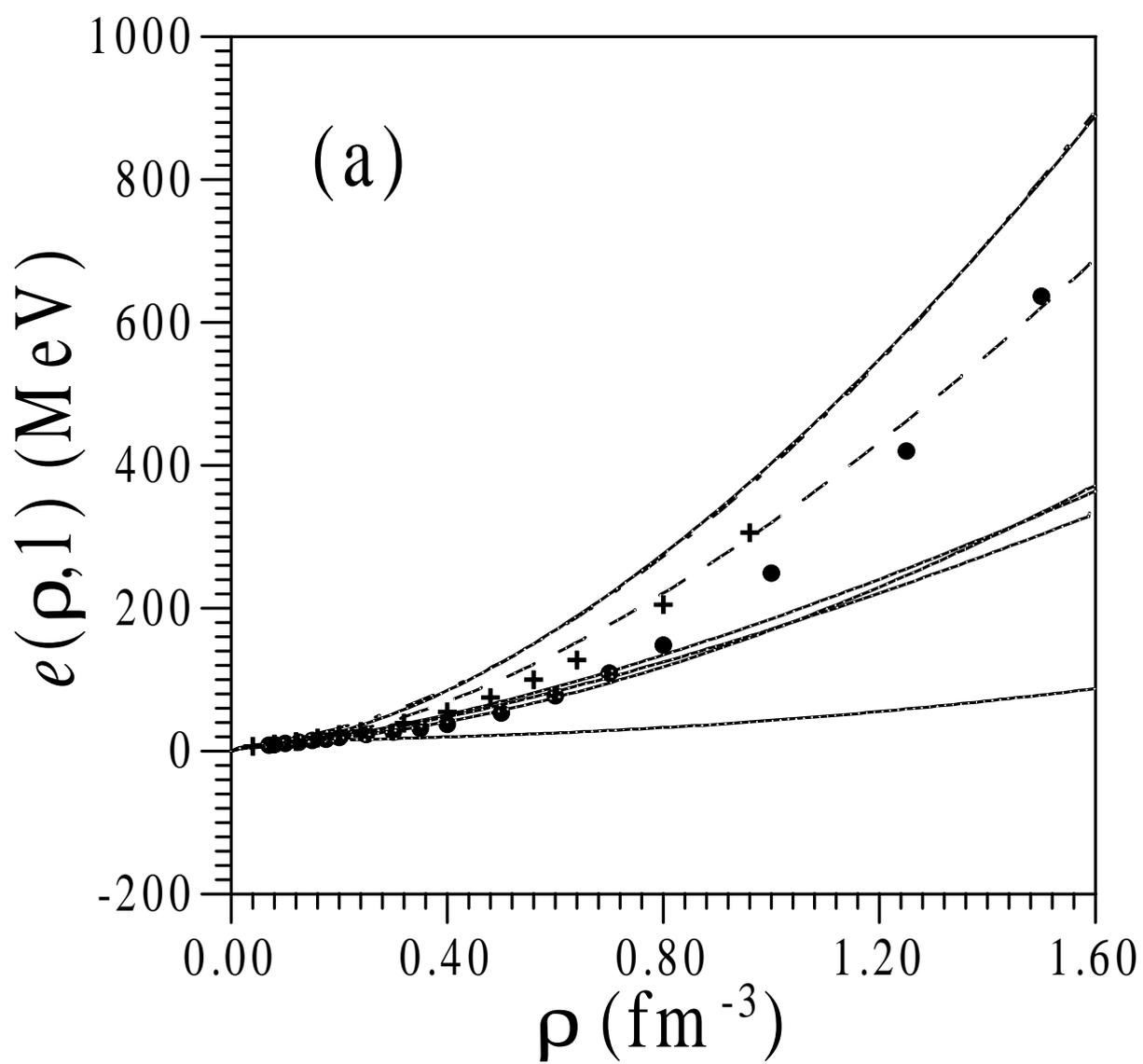
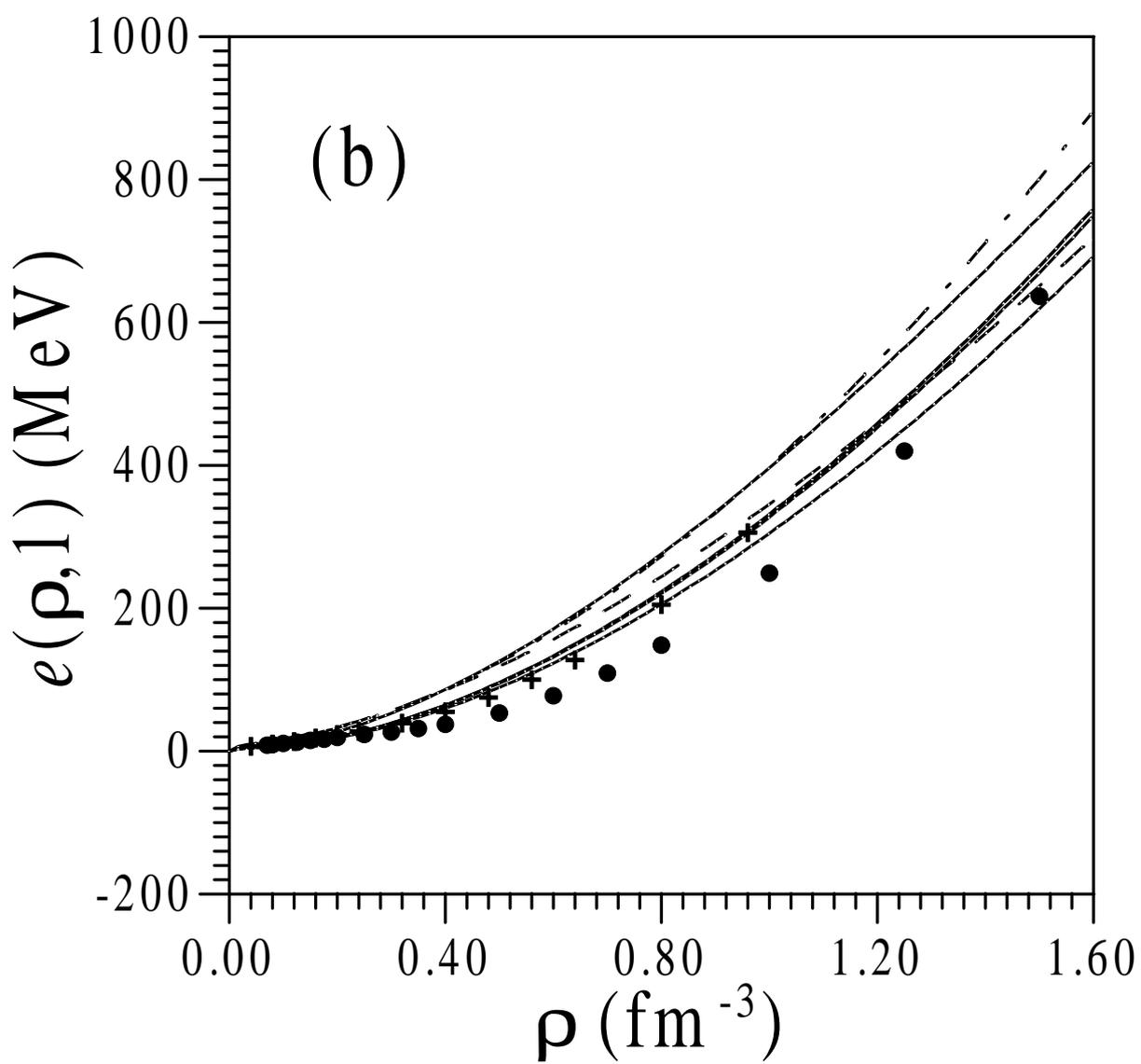
\end{document}